\documentclass[11pt]{article}

\usepackage{jheppub}

\usepackage[utf8]{inputenc}
\usepackage{setspace}
\usepackage{verbatim}
\usepackage[usenames,dvipsnames]{xcolor}

\usepackage{amsmath}
\usepackage{latexsym}
\usepackage{tikz}

\usepackage{mathtools}

\usepackage{float}
\usepackage{bm}

\renewcommand{\d}{\mathrm{d}}

\newcommand{\be}{\begin{equation}}

\newcommand{\ket}[1]{| #1 \rangle}

\makeatletter
\newcommand{\subalign}[1]{%
  \vcenter{%
    \Let@ \restore@math@cr \default@tag
    \baselineskip\fontdimen10 \scriptfont\tw@
    \advance\baselineskip\fontdimen12 \scriptfont\tw@
    \lineskip\thr@@\fontdimen8 \scriptfont\thr@@
    \lineskiplimit\lineskip
    \ialign{\hfil$\m@th\scriptstyle##$&$\m@th\scriptstyle{}##$\crcr
      #1\crcr
    }%
  }
}
\makeatother

\usepackage{amssymb}
\usepackage{graphicx}
\usepackage{svg}

    \newcommand{\beq}{\begin{equation}}
    \newcommand{\eeq}{\end{equation}}
    \newcommand\beqa{\begin{eqnarray}}
    \newcommand\eeqa{\end{eqnarray}}
\def\<{\left<}
\def\>{\right>}
\def\d{\partial}
\newcommand{\nn}{\nonumber}
\newcommand{\eq}[1]{(\ref{#1})}

\title{Integrable Feynman Graphs and Yangian Symmetry on the Loom}

\emailAdd{kazakov$\bullet$lpt.ens.fr}
\emailAdd{fedor.levkovich$\bullet$gmail.com}
\emailAdd{mishnyakovvv$\bullet$gmail.com}

\author[a]{Vladimir Kazakov,}
\author[b]{Fedor Levkovich-Maslyuk,}
\author[c,d,e,f]{Victor Mishnyakov}

\affiliation[a]{
 Laboratoire de Physique de l'\'{E}cole Normale Sup\'{e}rieure,
 CNRS, Universit\'{e} PSL, Sorbonne Universit\'{e}s,
 24 rue Lhomond, 75005 Paris, France
 }

 \affiliation[b]{
 Universit\'{e} Paris Saclay, CNRS,  CEA, Institut de physique th\'{e}orique,   91191, Gif-sur-Yvette, France
 } 

 \affiliation[c]{  Lebedev Physics Institute, Moscow 119991, Russia
 } 
 \affiliation[d]{  NRC ``Kurchatov Institute'', 123182, Moscow, Russia
 } 
 \affiliation[e]{ MIPT, Dolgoprudny 141701, Russia
 } 
 \affiliation[f]{Institute for Theoretical and Mathematical Physics, Lomonosov Moscow State University, Moscow 119991, Russia}

\abstract{
 We present significant evidence that the powerful property of Yangian invariance extends to a new large class of conformally invariant Feynman integrals. Our results apply to planar Feynman diagrams in any spacetime dimension dual to an arbitrary network of intersecting straight lines on a plane (Baxter lattice), with propagator powers determined by the geometry. We  formulate Yangian symmetry in terms of a chain of Lax operators acting on the fixed coordinates around the graph, and we also extend this construction to the case of infinite-dimensional auxiliary space. Yangian invariance leads  to new differential and integral equations for individual, highly nontrivial, Feynman graphs, and we present them explicitly for several examples.  The graphs we consider determine correlators in the recently proposed loom fishnet CFTs. We also describe a generalization to the case with interaction vertices inside open faces of the diagram. 
 Our construction unifies   and greatly extends the known special cases of Yangian invariance  to likely the most general family of integrable scalar planar graphs.

}

\begin{document}

\maketitle

\section{Introduction}

Yangian symmetry for quantum integrable systems is an old and well established subject~\cite{Faddeev:1996iy,molev2007yangians,Bernard:1992ya,MacKay:2004tc,Loebbert:2016cdm}.
In physics, besides spin chains, one of the most important uses of the Yangian symmetry concerns  the planar scattering amplitudes in  $N=4$ supersymmetric Yang-Mills theory or in ABJM theory where it is closely linked with dual conformal symmetry~\cite{Drummond:2009fd,Drummond:2008vq,Beisert:2010gn,Arkani-Hamed:2012zlh,Huang:2010qy,Bargheer:2010hn} \footnote{See also \cite{Beisert:2017pnr} for another related set of applications.}. However its application to planar conformal Feynman graphs  is a rather recent observation~\cite{Chicherin:2017cns,Chicherin:2017frs} (see also \cite{Chicherin:2013sqa}).\footnote{Such integrable planar graphs defined in position space have an important advantage w.r.t. the amplitudes: unlike the latter they are finite objects requiring no IR/UV regulators, so that Yangian symmetry manifests itself as a rigorous mathematical statement.} As a powerful and rather general tool, Yangian symmetry  can have far-reaching consequences, both for the formal development  of  quantum integrability as well as for practical calculation of interesting physical quantities: specific Feynman graphs and planar scattering amplitudes, partition functions of integrable statistical mechanical systems with spins fixed at arbitrary boundaries, as well as for the analysis of moduli space of Calabi-Yau manifolds~\cite{Duhr:2022pch}. 
%Yangian symmetry lies among a diverse variety of properties of Feynman diagrams and specifically the different types of differential equations satisifed by Feynman diagrams 
Some new developments stemming from these implementations of Yangian symmetry can be found in~\cite{Loebbert:2019vcj,
Loebbert:2020hxk,Loebbert:2020tje,Loebbert:2020glj,Corcoran:2020epz,Corcoran:2021gda}, see \cite{Loebbert:2022nfu,Chicherin:2022nqq} for a review. This symmetry is expected to be of particular importance for the fishnet CFTs, first discovered in~\cite{Gurdogan:2015csr} as a special scaling limit of the $\gamma$-deformed $\mathcal{N}=4$ Super-Yang-Mills theory (see ~\cite{Caetano:2016ydc,Kazakov:2018hrh} for a review).  The fishnet theories were later generalized to the fishnet reduction of 3d ABJM theorys~\cite{Caetano:2016ydc} and to other spacetime dimensions and deformations~\cite{Kazakov:2018qbr,Mamroud:2017uyz}  and finally to the most general setup -- the loom fishnet CFTs~\cite{Kazakov:2022dbd}. The graph content of these fishnet CFTs is  based on the old construction of A.~Zamolodchikov~\cite{Zamolodchikov:1980mb}   for the most general integrable planar conformal Feynman diagrams, which he considered as a specific statistical mechanical model. Based on this integrability of the underlying Feynman graphs, many physical results have been found for these fishnet CFTs, for various quantities and in various dimensions~\cite{Derkachov:2021ufp,Derkachov:2021rrf,Derkachov:2020zvv,Derkachov:2019tzo,Kazakov:2018gcy,Derkachov:2018rot,Kazakov:2018qbr,Caetano:2016ydc,Gromov:2018hut,Gromov:2019aku,Gromov:2017cja,Grabner:2017pgm,Basso:2018cvy,Basso:2018agi,Basso:2017jwq,Mamroud:2017uyz,Cavaglia:2021mft,Pittelli:2019ceq,Kostov:2022vup,Ferrando:2023ogg}. 

Yangian symmetry gives a remarkable realization of integrability at the level of an individual Feynman graph. It has been used successfully in order to obtain a variety of new results for nontrivial conformal Feynman integrals, with  the first direction explored \cite{Chicherin:2017cns,Chicherin:2017frs} being arbitrary graphs in the original 4d fishnet theories where the graphs are cut out from a regular square lattice. These results were later generalized to specific examples in other dimensions and/or with generic propagator powers \cite{Loebbert:2019vcj,Corcoran:2021gda,Duhr:2022pch}. More generally, in \cite{Chicherin:2022nqq,Loebbert:2020hxk} it was suggested that conformal graphs in any dimensions cut out from regular tilings of the plane should have Yangian symmetry. At the same time, a unified picture of what is the full class of graphs with Yangian symmetry has been missing, and the exploration of this powerful property has been done mostly on a case-by-case basis.

% establish the Yangian symmetry of the most general planar Feynman diagrams

\begin{figure}[h]
 \centering
   \includegraphics[scale=0.8]{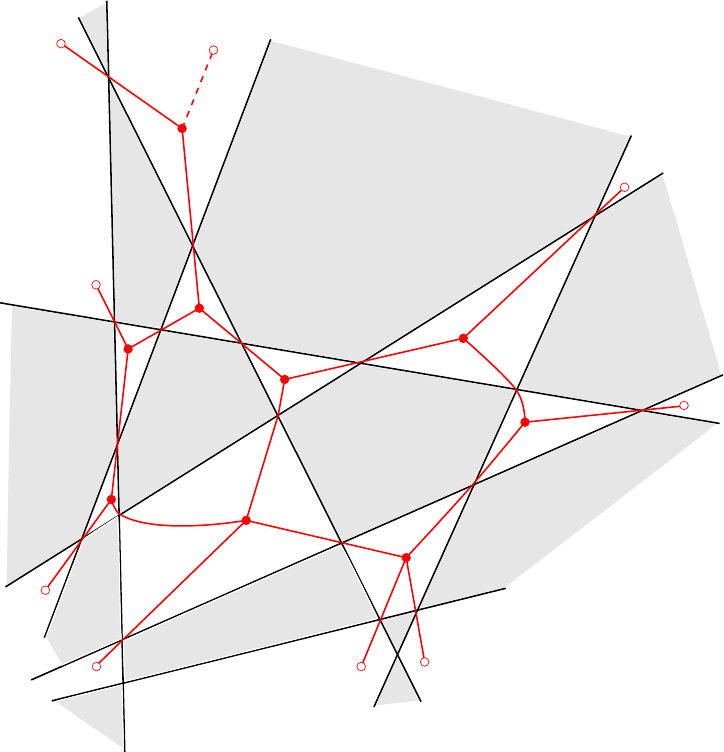}  \caption{Example of a Feynman graph generated via the loom construction. The propagators are shown as red lines, the internal vertices filled red circles, and the external points as empty red circles.  Black lines form the original Baxter lattice which is painted in the checkerboard manner -- the loom. To each internal vertex or external point we assign a $D$-dimensional coordinate and integrate over those of the internal points. At the top left we show one generalisation we have made to the standard construction, namely a vertex inside  an open external face with an extra dashed propagator (more details in section \ref{sec:gen}).}
   \label{fig:loomexnolass}
 \end{figure}

In this work, we will generalize
the lasso method of ~\cite{Chicherin:2017cns,Chicherin:2017frs} to a large new class of Feynman integrals, giving
a valuable tool to prove Yangian invariance of these integrals. While the results of~\cite{Chicherin:2017cns,Chicherin:2017frs} are for the particular case of 4d planar graphs of disc topology cut out of the regular square lattice, we will study the case of graphs built via Zamolodchikov’s general integrability construction based on star-triangle relations. These graphs are defined starting from a general (and not necessarily regular) lattice dual to the so-called Baxter lattice \cite{baxter1978solvable}\footnote{in \cite{baxter1978solvable} it was originally called `Baxter Z-invariant lattice'}  -- an arbitrary collection of intersecting straight lines that provide a checkerboard coloring of the plane,
see figure \ref{fig:loomexnolass}\ \footnote{We will often employ the name ``loom" introduced in~\cite{Kazakov:2022dbd} for such a lattice, indicating that it is a device for ``weaving"  integrable planar graphs}. The scaling dimensions of the propagators are expressed via the angles at which the lines intersect. This geometric construction ensures that the graphs are finite and conformal (i.e. the sum of scaling dimensions at each vertex is $D$), and in addition imposes further linear constraints on the propagator powers so that not all conformal graphs are integrable in the sense we discuss\footnote{see a discussion of these constraints in section 2 of~\cite{Kazakov:2022dbd}}.  As an outcome, the graph is shown to be an eigenstate of the 'lasso' monodromy matrix constructed as a chain of conformal Lax operators known from \cite{Chicherin:2012yn}.  

In order to demonstrate Yangian invariance, we gradually remove the lasso, i.e. the chain of Lax operators, from the Feynman diagram step by step, extending the approach of  \cite{Chicherin:2017cns,Chicherin:2017frs}. In the process we use a variety of nontrivial identities for the conformal Lax operators. We have managed to rigorously prove the validity of several steps in this (in general quite complex) procedure by invoking a number of geometric arguments based on the way the graph is constructed from the Baxter lattice. We expect that the procedure should apply to an arbitrary Feynman graph constructed in this way.

We also further extend the class of admissible Feynman graphs by generalising the construction of \cite{Zamolodchikov:1980mb,Kazakov:2022dbd} -- namely, allowing the interaction vertices to be placed inside external open faces of the lattice. We give more details on this in section  \ref{sec:gen}.

As a main result, we will derive, expanding in inverse powers of the spectral parameter, the linear Yangian differential  equations on these graphs. The equations have a pretty universal form and are distinguished only by different sets of evaluation parameters for different graphs. Such equations have been obtained before for various particular graphs, and in many cases were shown to be rather powerful as they restrict the graph to be a linear combination of a finite set of basis functions. This has already led to the calculation of several graphs including first the general cross and then the double cross for which the result is highly involved and was found only recently \cite{Loebbert:2019vcj,Ananthanarayan:2020ncn}. We hope that our results will open the way to exploration of more general Feynman integrals.

Furthermore, we managed to construct the lasso operator, i.e. the chain of Lax matrices, with a noncompact representation in the auxiliary space (instead of the compact one mentioned above), the same as the principal series representation of $D$-dimensional conformal algebra in the quantum space. The corresponding R-matrix out of which our lasso  is made has been known for quite a while~\cite{Chicherin:2012yn}\footnote{following the earlier approaches  of \cite{Derkachov:2001yn,Derkachov:2002wz,Derkachov:2002tf} for $sl_2$ integrability}.  The statement that the graph is an eigenfunction of this lasso of R-matrices is now an \textit{integral} rather than a \textit{differential} equation. Furthermore, the R-matrix itself is simply a product of four propagators with weights containing the spectral parameter, which means that our construction not only gives new integral equations for the graph but also represents an interesting equivalence  between various Feynman diagrams.

The  paper is organized as follows. In section \ref{sec:gr} we review Zamolodchikov's construction of Feynman graphs based on the loom -- the Baxter lattice  -- and describe our generalisation of it. In section \ref{sec:lax} we review the conformal Lax operators and their key properties. In section \ref{sec:yang} we describe the lasso construction which leads to Yangian invariance of the graphs and its extension to our general case, with a rather intricate proof of several steps in the process of removing the chain of Lax operators from the diagram. In section \ref{sec:diff} we present the general form of the differential equations for Feynman integrals which follow from Yangian invariance, as well as some explicit examples. In section \ref{sec:infd} we describe another nontrivial generalisation of the construction, namely the case of infinite-dimensional auxiliary space and the resulting integral equations for graphs. We present conclusions and future directions in section \ref{sec:concl}.

\section{Integrable Feynman graphs from the loom}
\label{sec:gr}

The Feynman graphs we discuss in this paper are coordinate-space planar diagrams   obtained from the construction originally proposed by Zamolodchikov in \cite{Zamolodchikov:1980mb}. Let us first discuss here this construction in detail, for completeness\footnote{See also its description in~\cite{Kazakov:2022dbd}}, and then in section \ref{sec:gen} we will describe a certain  generalisation of it that we will also explore.

\begin{figure}[h]
 \centering
   \includegraphics[scale=0.3]{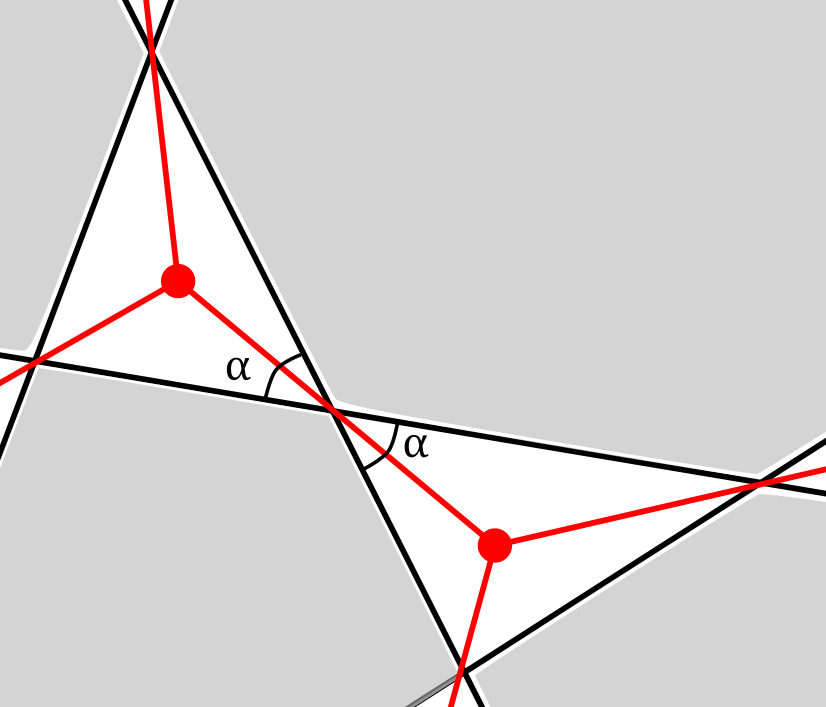}  \caption{Fragment of a Feynman graph. The propagator between points $x_1,x_2$ is given by $(x_1-x_2)^{-2\Delta}$ with the power determined by the angle $\alpha$ through which it passes according to $\Delta=D\frac{\pi-\alpha}{2\pi}$.}
   \label{fig:propdem1}
 \end{figure}

The starting point is a Baxter lattice -- a finite set of intersecting lines on the plane. Some lines may be parallel to each other but  triple or higher intersections are not allowed. Such a set of lines divides the plane into a set of polygonal faces which admit a checkerboard coloring, and accordingly we will draw them as white or grey. The actual Feynman diagram is drawn on the graph which is \textit{dual}\footnote{i.e. its vertices correspond to faces of the original graph, and vice versa} to the lattice of the white faces. The rules are as follows:
 \begin{itemize}
     \item Inside any white face which is a closed polygon  one may place an internal vertex of the Feynman graph. In this case we draw propagators going from this vertex to each of its neighboring white faces   and passing through the angles of this polygon.  To each internal vertex we associate a $D$-dimensional coordinate which is to be integrated over. 
     \item A propagator coming out of an internal vertex can be either an external leg of the diagram, or it can connect to the internal vertex in the neighboring white face.
     \item The propagator connecting two points $x_1$ and $x_2$ is given by the conformal 2-pt function
   \beq
   \label{cprop}
     \frac{1}{|x_{1}-x_{2}|^{2\Delta}}\ ,
   \eeq
    and, crucially, the scaling dimension $\Delta$ is determined by the angle of the polygon through which the propagator passes as (see figure \ref{fig:propdem1})
   \beq
   \label{da}
    \Delta=D\frac{\pi-\alpha}{2\pi} \ .
 \eeq
 \end{itemize}
As a result we obtain a Feynman diagram given by a multipoint  integral in coordinate space with propagators of the form \eq{cprop}.

Remarkably, the sum of scaling dimensions for propagators at each vertex is $D$ due to the  geometric constraint on the sum of the angles of any closed polygon. To see this, notice that $\pi -\alpha$ in \eq{da} is the external angle at the vertex of the polygon, and the sum of all such angles for an $n$-gon is $\pi n - \pi(n-2)=2\pi$ (regardless of $n$). Thus the resulting Feynman integral is always a conformal object. We show an example of such a Feynman graph on figure \ref{fig:loomexnolass} in the Introduction.

 \begin{figure}[h]
 \centering
   \includegraphics[scale=1.4]{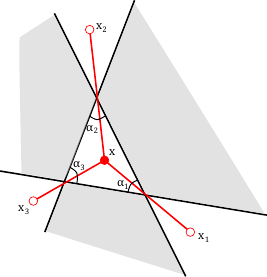}  \caption{A simple Feynman graph with one internal vertex and three external points.}
   \label{fig:triang0}
 \end{figure}

The simplest example is a 1-point integral with three external legs, shown on figure \ref{fig:triang0}. It is given by
\beq
    I=\int d^Dx \frac{1}{|x_1-x|^{2\Delta_1}|x_2-x|^{2\Delta_2}|x_3-x|^{2\Delta_3}} \ ,
\eeq
where $\Delta_k=D\frac{\pi-\alpha_k}{2\pi}$ are fixed in terms of the three angles of the triangle. This integral can be computed via the star-triangle relation that we review below (equation \eq{str}).

So far we have defined these Feynman graphs just as explicit multipoint integrals, without regard to any particular field theory. However, it was recently understood in \cite{Kazakov:2022dbd} that there does exist a large family of Lagrangian field theories, dubbed as loom fishnet CFTs, whose Feynman diagrams  that determine perturbative correlation functions of single trace operators are precisely these ones. This further motivates their deep investigation.

 \begin{figure}[h]
 \centering
   \includegraphics[scale=1]{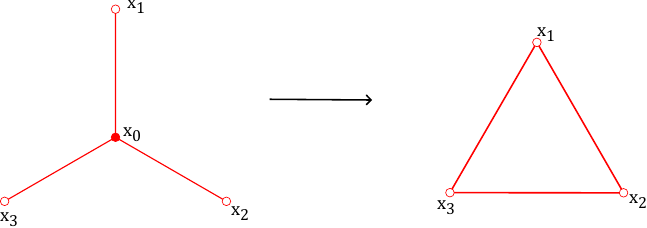}  \caption{Graphic illustration of the star-triangle relation \eq{str}.}
   \label{fig:str}
 \end{figure}

When working with these Feynman graphs it is often useful to utilise the star-triangle identity. It  reads
\beq
\label{str}
    \int d^Dx_0\prod_{i=1}^3\frac{1}{|x_{i0}|^{2\Delta_i}}=\pi^{D/2}\prod_{i=1}^3\frac{\Gamma(D/2-\Delta_i)}{\Gamma(\Delta_i)}\prod_{i=1}^3\frac{1}{|x_{i-1}-x_{i+1}|^{2(D/2-\Delta_i)}} \ ,
\eeq
and holds when $\Delta_1+\Delta_2+\Delta_3=D$. We show it graphically on figure \ref{fig:str}. Let us also mention that this identity has a very natural place in the setting we are considering since moving the lines of the Baxter lattice amounts to star-triangle transformations of the Feynman graph \cite{Zamolodchikov:1980mb}. This fact itself may be viewed as a manifestation of the `integrability' of these graphs.

\subsection{Generalization to vertices inside external faces}
\label{sec:gen}

\begin{figure}[h]
 \centering
   \includegraphics[scale=0.7]{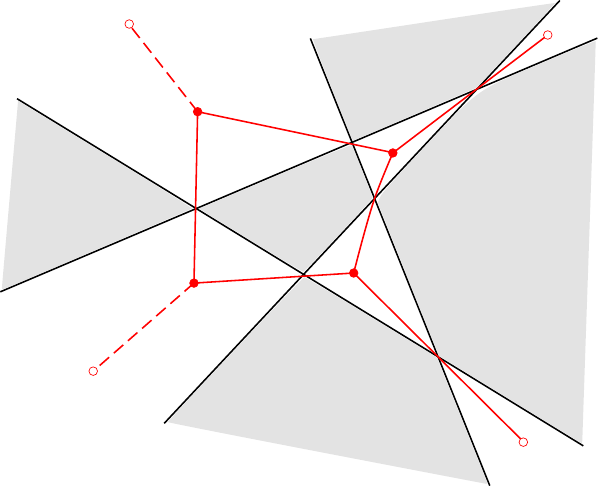}  \caption{A generalized loom Feynman diagram with interaction vertices inside external faces. The Feynman graph is built out of propagators shown in red. Bold filled circles show internal integrated vertices, empty circles show the ends of external legs. The scaling dimensions for dashed lines are fixed by requiring the sum of dimensions at each vertex to be $D$, for the other lines they are fixed by geometry (see section \ref{sec:gr}).}
   \label{fig:sqex1}
 \end{figure}

\begin{figure}[h]
 \centering
   \includegraphics[scale=1]{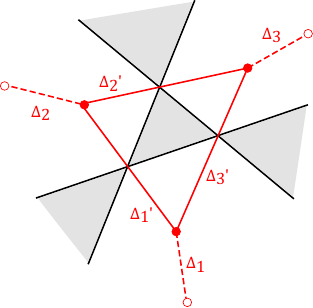}  \caption{A simple Feynman graph with interaction vertices inside external open faces. We have labelled in red the scaling dimensions for each propagator.}
   \label{fig:triang1}
 \end{figure}

In the construction we described above, internal vertices are only allowed inside faces that are \textit{closed} polygons, and not inside the external open faces that go off to infinity. This is natural since the fixed sum of angles of a closed polygon guarantees that the sum of  weights at each vertex is $D$ and the graph is conformal. In fact we found that this construction can be naturally extended also to the case when internal vertices are placed inside \textit{open} faces. In this case one should simply add an extra propagator which completes the sum of conformal weights at the vertex to $D$. We draw these propagators as dashed lines, examples are given on figures \ref{fig:sqex1}, \ref{fig:triang1}. Another example is given in the Introduction on figure \ref{fig:loomexnolass} where a vertex of the new type is in the upper left corner.

Notice that the new propagator will always have a positive scaling dimension as follows from geometry. That is, when we remove some edges from a closed polygon the sum of its external angles will become smaller than $2\pi$ which in terms of scaling dimensions means the sum of original $\Delta$'s is less than $D$, so the new propagator we add will bring the sum up to $D$.

Notice that both the triangle graph from figure \ref{fig:triang1} it and the square graph from figure \ref{fig:sqex1} \textit{cannot} be drawn on a standard loom unless we allow these extra vertices, the reason being that they have `too few' external legs (this is not hard to prove rigorously). Both of these examples in the end can be reduced to simpler integrals with only one or two internal vertices by applying the star-triangle relation\footnote{Concretely, for the square one can convert the two opposite vertices to triangles. For the triangle, one can e.g. convert the triangle formed by internal lines into a star.}, however they illustrate the idea that some graphs can only be drawn using this new type of vertices. In would be interesting to understand precisely which graphs can be drawn using these new vertices in addition to the standard loom construction.

In the remainder of the paper we will work with Feynman graphs built using the procedure we just described and we will show how Yangian invariance can be implemented for them.

\subsection{Comments on the construction of Feynman graphs}\label{sec:comments}

Let us discuss several important properties and immediate consequences of the construction of Feynman graphs we  presented above. While some of these remarks were mentioned in \cite{Kazakov:2022dbd}, we find it useful to summarize them and others in a systematic way here. We will illustrate these points with examples which, while sometimes very simple, still highlight the feature in question that by itself can appear in much more complicated situations.

\begin{figure}[h]
 \centering
   \includegraphics[scale=0.9]{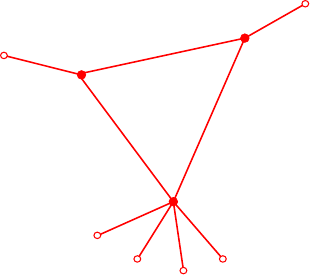}  \caption{A Feynman graph that cannot be drawn on a loom.}
   \label{fig:trmany}
 \end{figure}

First, notice that even when we allow for vertices to be placed inside open faces, not any planar Feynman graph can be even drawn on a loom. An example is the triangle graph shown on figure \ref{fig:trmany}, where we have many external legs coming out of one of the vertices and only one leg for each of the remaining two vertices. The reason is that, starting from a triangle with only three external legs (figure \ref{fig:triang1}), in order to add extra legs to a vertex one needs to add extra lines to the Baxter lattice, which will necessarily create more intersections corresponding to external legs at other vertices as one can easily see.

Second, even when the topology of a graph is compatible with obtaining it from the loom, the loom construction implies additional relations between scaling dimensions of the propagators besides those that ensure conformality, i.e. besides the relation that the sum of dimensions at each vertex is $D$. In other words, not all conformal graphs are in the integrable class (see \cite{Kazakov:2022dbd} for a recent discussion of this).  To illustrate the existence of these relations, consider the 3-point graph shown on figure \ref{fig:triang1}. Here in addition to the three constraints coming from the sum of dimensions being $D$ at each vertex,
\beq
    \Delta_1+\Delta_1'+\Delta_3'=D \ , \ \  \Delta_2+\Delta_1'+\Delta_2'=D \ , \ \  \Delta_3+\Delta_2'+\Delta_3'=D \ , \ 
\eeq
we have an extra 'non-local' (i.e. not associated to a single internal vertex) constraint
\beq
    \Delta_1'+\Delta_2'+\Delta_3'=D/2 \ .
\eeq
It follows from expressing the dimensions in terms of the angles $\alpha_i$ ($i=1,2,3$) of the grey inner triangle on figure \ref{fig:triang1} as $\Delta_i'=D/(2\pi)\alpha_i$  and using $\alpha_1+\alpha_2+\alpha_3=\pi$. Conversely, this constraint allows for the application of star-triangle relation to the internal triangle.  These extra relations between scaling dimensions are  often crucial for ensuring  integrability properties such as Yangian symmetry as we will see later on. More involved examples of these non-local constraints are given in section \ref{sec:diff}.

 \begin{figure}[h]
 \centering
   \includegraphics[scale=0.8]{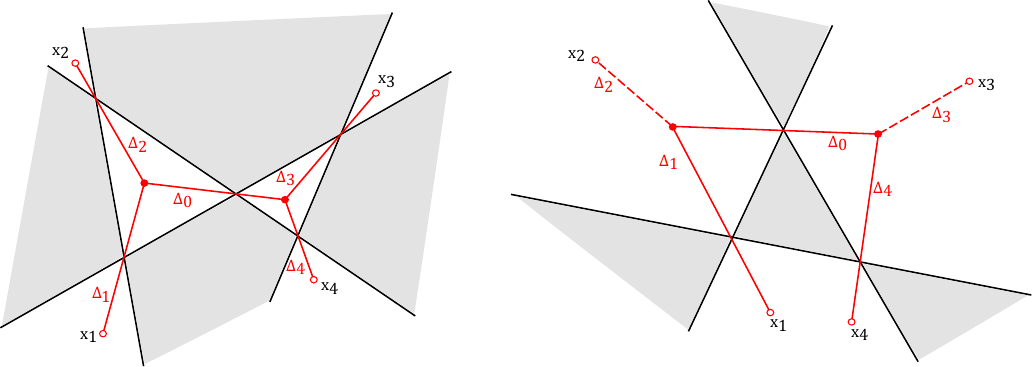}  \caption{The same Feynman graph drawn on two different looms.}
\label{fig:kitespl}
 \end{figure}

Third, somewhat surprisingly, it can happen that the \textit{same} Feynman graph can be drawn on two \textit{different} looms. In both cases the result has Yangian symmetry, but the two constructions can lead to two different sets of constraints on its scaling dimensions. In particular it could happen that one set of constraints is weaker than the other one but is still sufficient for integrability. An example is shown on figure \ref{fig:kitespl}. In this case the left figure gives constraints
\beq
\Delta_1+\Delta_2+\Delta_0=D \ , \ \ \ \Delta_3+\Delta_4+\Delta_0=D \ ,
\eeq
while the right one gives an \textit{additional} constraint
\beq
\label{nl1}
    \Delta_1+\Delta_0+\Delta_4=D/2 \ ,
\eeq
which follows from considering the angles of the grey triangle in the middle. 
As the first representation already guarantees Yangian symmetry for the graph, it is clear that this new constraint \eq{nl1} is in fact not needed for it\footnote{One can also check this statement explicitly using the techniques from section \ref{sec:yang} step by step.}. Thus in this case one representation provides a strictly weaker set of constraints that ensure integrability. Accordingly, when investigating whether or not a particular graph is in the integrable class one should try to look for different ways to obtain it from the loom\footnote{In principle it might  also happen that two representations lead to two different sets of constraints, neither of which is strictly stronger than the other one, leading to two different integrable loci in the space of the graph's parameters.}. It would be interesting to find an algorithm that provides the minimal set of constraints for a given graph.

    Fourth, notice that it could happen that two external legs end inside the same face, see figure \ref{fig:kitespl} (right) for an example. While in general the logic of the construction of \cite{Zamolodchikov:1980mb} suggests there is only one $x$ variable to be associated with each face, in this case we observed that we do have Yangian symmetry if we keep the external coordinates for these legs distinct (on figure \ref{fig:kitespl} these are $x_1\neq x_4$). This seems quite natural since in general bringing together coordinates of some external legs is also a potentially dangerous procedure with which the Yangian symmetry may not commute (see e.g. \cite{Loebbert:2020tje,Corcoran:2021gda}).

\section{Conformal algebra and Lax operators}
\label{sec:lax}

We work in the conventions of \cite{Chicherin:2012yn} and we can take the metric to be either Minkowski,\footnote{The computations we discuss are equivalently valid in either Euclidean or Minkowski signature. The final result, i.e. differential equations for Feynman integrals, has the same form in both cases, but the analytic continuation of the integral from one to the other can be subtle, see \cite{Corcoran:2020epz,Chicherin:2017frs} for more details}
\beq
    g_{\mu\nu}={\rm diag}(1,-1,-1,\dots,-1) \ , \ \ \ \ \ \mu,\nu=0,\dots,D-1 \ ,
\eeq
or Euclidean.
The conformal algebra generators acting on scalar fields of dimension $\Delta$ have the form
\beq
\label{conf1}
    P_\mu=-i\d_{x_\mu} \equiv\hat p_\mu\ , \ \ D=x^\mu \hat p_\mu-i\Delta \ ,
\eeq
\beq
\label{conf2}
    L_{\mu\nu}=x_\nu\hat p_\mu-x_\mu\hat p_\nu \equiv \hat\ell_{\mu\nu}\ , \ \ \  K_\mu=2x^\nu\hat\ell_{\nu\mu}+x_\nu x^\nu\hat p_\mu-2i\Delta x_\mu \ .
\eeq
We also define
\beq
   {\bf x}=-i{\bm{\bar\sigma}}^\mu x_\mu \ , \ \ \  {\bf p}=-\frac{i}{2}{\bm \sigma}^\mu\d_{x_\mu} \ ,
\eeq
using the sigma matrices which in arbitrary even dimension $D$ have size $2^{D/2-1} \times 2^{D/2-1}$ (see \cite{Chicherin:2012yn} for details).

The Lax operator for the conformal group reads \cite{Chicherin:2012yn}
\beq
\label{lax}
    L_{\alpha\beta}(u_+,u_-)=\begin{pmatrix}
    u_+-{\bf p}{\bf x} & {\bf p} \\ 
    {\bf x}(u_+-u_-)-{\bf x}{\bf p}{\bf x} & {\bf x}{\bf p}+u_-
    \end{pmatrix} \ .
\eeq
It is a matrix of size $2^{D/2}\times 2^{D/2}$ with indices taking values $\alpha,\beta=1,\dots,2^{D/2}$. Here we took the physical space to be the scalar representation with dimension $\Delta$ while the auxiliary space is the $2^{D/2}$-dimensional spinor representation of the conformal group. The parameters $u_\pm$ encode the spectral parameter $u$ and the dimension $\Delta$ via
\beq
    u_+=u+\frac{\Delta-D}{2} \ , \ \ u_-=u-\frac{\Delta}{2} \ ,
\eeq
which implies
\beq
\label{du}
    \Delta=u_+-u_-+D/2 \ , \ \ u=\frac{1}{2}(u_++u_-+D/2) \ .
\eeq
Below we will use shortened notation for the Lax operator with shifted arguments,
\beq
   \label{sqb}
    L[\delta^+,\delta^-]\equiv L[u+\delta^+,u+\delta^-] \ ,
\eeq
and similarly for a single number
\beq
    [\delta]\equiv u+\delta \ .
\eeq

The above construction of the Lax operator is formulated for even dimension only. However, its final outcome are differential equations for the graphs discussed in section \ref{sec:diff}, in which the dimension appears just as a parameter. It is natural to expect that these equations should hold in any dimension (including odd $D$).

\subsection{Properties of Lax operators}

Below we will extensively use several important properties of this Lax operator. First, we have the key intertwining relation (derived in \cite{Chicherin:2012yn}, see equation (5.4) there)
\beq
\label{intw}
    L_1(u+\Delta,u')L_2(v,u)\frac{1}{x_{12}^{2\Delta}}=\frac{1}{x_{12}^{2\Delta}}L_1(u,u')L_2(v,u+\Delta) \ ,
\eeq
which allows one to move a propagator through two Lax operators with appropriately chosen arguments. Furthermore, for specially chosen arguments $L$ acts diagonally on a constant function,
\beq
\label{laxon1}
    L_{\alpha\beta}(u,u+D/2)\cdot 1=(u+D/2)\delta_{\alpha\beta} \ ,
\eeq
as one can verify using identities between sigma matrices given in \cite{Chicherin:2012yn}. 
We will also denote by $L^T$ the Lax operator obtained from the original one by a `transposition' in the physical space\footnote{this is unrelated to a transposition in the $\alpha,\beta$ indices in auxiliary space} which amounts to integration by parts, i.e. replacing $x^\mu\to x^\mu,\;\d_\mu\to -\d_\mu$ and reversing the order of all $x$ and $\d_x$ operators. In other words, we have
\beq
    \int d^D x\; g(x)L_{\alpha\beta}(u,v)f(x)=\int d^D x\; (L_{\alpha\beta}^T(u,v)g(x))f(x) \ .
\eeq
Then one can verify that similarly to \eq{laxon1} we have
\beq
\label{lte}
    L^T_{\alpha\beta}(u+D/2,u)\cdot 1=(u+D/2)\delta_{\alpha\beta} \ .
\eeq
We will also use the inversion formulas which can be checked by direct computation and generalize those from \cite{Chicherin:2017cns} to any (even) dimension $D$,
\beq\label{inv1}
    L(u_+,u_-)^{-1}=-\frac{1}{u_+u_-}L(-u_-,-u_+) \ ,
\eeq
\beq\label{inv2}
    (L(u_+,u_-)^t)^{-1}=-\frac{1}{(u_++D/2)(u_-+D/2)}L^t(-u_--D,-u_+-D) \ ,
\eeq
where $t$ denotes transposition in the matrix indices.

\section{Yangian invariance}
\label{sec:yang}

\begin{figure}[h]
 \centering
   \includegraphics[scale=0.75]{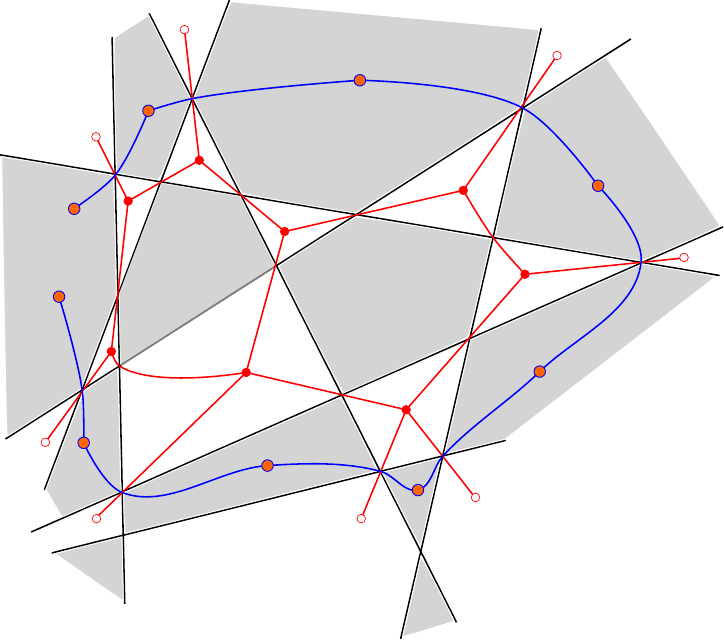}  \caption{The lasso and the Feynman graph on the loom. The chain of Lax operators is shown in blue, with the orange circles showing contraction of the indices of adjacent Lax operators.}
   \label{fig:lass1}
 \end{figure}

In this section we show that the Feynman graphs obtained from the loom construction are Yangian invariants. Namely, they are eigenstates of the monodromy matrix built as a product of $n$ conformal Lax operators \eq{lax} where $n$ is the number of external legs of the graph,
\beq
    (L_n[\delta_n^+,\delta_n^-]\dots L_2[\delta_2^+,\delta_2^-]L_1[\delta_1^+,\delta_1^-])_{\alpha\beta} \ket{G}=\delta_{\alpha\beta}\lambda(u)\ket{G} \ .
    \label{gre}
\eeq
Here we denoted by $\ket{G}$ the Feynman integral corresponding to the graph, and the Lax operator $L_k$ acts on the coordinate $x_k$ of the $k$-th external leg of the graph (we used the notation \eq{sqb} for arguments of Lax operators). Notice that the graph is an eigenstate of all the monodromy matrix elements (not just the trace as usual for spin chains). The shifts $\delta_k^\pm$ are read off from the geometry of the graph as we describe in detail below. We illustrate how the chain of Lax operators acts on the graph on figure \ref{fig:lass1}.

\subsection{Simple example: cross integral with generic dimensions}
\label{sec:cross}

To illustrate how the construction works let us start with a simple  example of a cross integral with generic dimensions $\Delta_k$ of the propagators,
\beq
\label{i41}
    I=\int \frac{d^Dx_0}{(x_{10})^{2\Delta_1}(x_{20})^{2\Delta_2}(x_{30})^{2\Delta_3}(x_{40})^{2\Delta_4}} \ ,
\eeq
with the constraint
\beq
    \sum_{k=1}^4\Delta_k=D
\eeq
that ensures the graph is conformal. When $D=4$ and all dimensions are set to $\Delta_i=1$ the result is given by the Bloch-Wigner function,
\beq
    I=\frac{\phi(z,\bar z)}{x_{13}^2x_{24}^2} \ , \ \ \ 
\eeq
with
\beq
    z\bar z=\frac{x_{12}^2x_{34}^2}{x_{13}^2x_{24}^2} \ , \ \ \ (1-z)(1-\bar z)=\frac{x_{14}^2x_{23}^2}{x_{13}^2x_{24}^2}
\eeq
and
\beq
    \phi=\frac{\pi^2}{z-\bar z}(2{\rm Li}_2(z)-2{\rm Li}_2(\bar z)+\log\frac{1-z}{1-\bar z}\log (z\bar z)) \ .
\eeq
For the case of generic $D$ and $\Delta$'s the statement of Yangian invariance is known from \cite{Loebbert:2019vcj} and the integral itself was found in that work to be a combination of Appell hypergeometric functions \cite{Usyukina:1992jd, Boos:1990rg}. Here it will serve as a useful pedagogical example to illustrate the construction. We will show that it's an eigenstate of the monodromy matrix
\beqa
\label{M4}
    M&\equiv& L_4[D,\Delta_1+\Delta_2+\Delta_3+D/2]
    L_3[\Delta_1+\Delta_2+\Delta_3,\Delta_1+\Delta_2+D/2] \\ \nn
    &\times&
      L_2[\Delta_1+\Delta_2,\Delta_1+D/2]
    L_1[\Delta_1,D/2] \ .
\eeqa
The derivation closely follows \cite{Chicherin:2017cns}, extending it to this more general case. The main trick is to use \eq{lte} to write 
\beq
\label{lt11}
    1=\frac{1}{[D/2]}L_0^T[D/2,0]\cdot 1 
\eeq
where we introduced an extra Lax operator associated with $x_0$ and acting here on a constant with arguments chosen so that the result is proportional to the identity. We insert \eq{lt11} under the integral in \eq{i41} and integrate by parts which amounts to replacing $L_0^T$ by $L_0$ so we have
\beqa
\label{il0}
    I&=&\int d^Dx_0\frac{1}{(x_{10})^{2\Delta_1}(x_{20})^{2\Delta_2}(x_{30})^{2\Delta_3}(x_{40})^{2\Delta_4}}\left(\frac{L_0^T[D/2,0]\cdot 1}{[D/2]}\right)
    \\ \nn 
    &=&\frac{ 1}{[D/2]}\int d^Dx_0 L_0[D/2,0]\frac{1}{(x_{10})^{2\Delta_1}(x_{20})^{2\Delta_2}(x_{30})^{2\Delta_3}(x_{40})^{2\Delta_4}} \ .
\eeqa
Next we consider the action of the monodromy matrix \eq{M4} on this integral. The first Lax matrix to act on it is $L_1$, in addition to $L_0$ which is already present in \eq{il0}. Nicely, we chose the arguments of the Lax matrices in such a way that we can use the intertwining relation \eq{intw} to move the first propagator $\frac{1}{(x_{10})^{2\Delta_1}}$ through the product $L_1L_0$, namely
\beq
    L_1[\Delta_1,D/2]L_0[D/2,0]\frac{1}{(x_{10})^{2\Delta_1}}=\frac{1}{(x_{10})^{2\Delta_1}}L_1[0,D/2]L_0[D/2,\Delta_1] \ .
\eeq
After this, $L_1$ now acts on a function which is independent of $x_1$ and furthermore its arguments differ by $D/2$, so we can use its action \eq{laxon1} on the identity and we can simply replace it by its eigenvalue which gives a factor $[D/2]$.

At the next step we now have to act with the $L_2$ operator. Again we can use the intertwining relation to pull the propagator through it,
\beq
    L_2[\Delta_1+\Delta_2,\Delta_1+D/2]L_0[D/2,\Delta_1]\frac{1}{(x_{20})^{2\Delta_2}}=\frac{1}{(x_{20})^{2\Delta_2}}L_2[\Delta_1,\Delta_1+D/2]L_0[D/2,\Delta_1+\Delta_2] \ .
\eeq
As before, we see that in the r.h.s. $L_2$ now acts on the constant function and can replaced by its eigenvalue $[\Delta_1+D/2]$. Repeating the same steps for $L_3$ and $L_4$, we see that again we can pull the corresponding propagators  through each one of them and in the end these Lax'es are replaced by their eigenvalues. One can see that at the very last step we will be left with action of $L_0$ on a constant with the arguments chosen as $L_0[D/2,\Delta_1+\Delta_2+\Delta_3+\Delta_4]$. Using that the sum of dimensions is $D$ we see that this Lax matrix becomes diagonal as well and just produces an extra scalar factor $[D]$. Collecting all the steps together, we find that the integral is indeed a Yangian invariant,
\beq
    M_{\alpha\beta}I=\lambda(u)\delta_{\alpha\beta}I \ ,
\eeq
and the eigenvalue reads
\beq
    \lambda(u)=[\Delta_1+D/2][\Delta_1+\Delta_2+D/2][\Delta_1+\Delta_2+\Delta_3+D/2][D] \ .
\eeq

This simple example illustrates the main idea one can use to prove Yangian invariance for more general graphs. Namely, we insert extra Lax operators at the internal vertices and then use a chain of intertwining relation to gradually get rid of Lax operators acting on parts of the graph. In \cite{Chicherin:2017cns} this was called the `lasso' method and we found that it extends to the more general graphs we consider here. We give a detailed derivation of how it works in the following subsections.

 A key part of the construction is choosing the labels $\delta_k^\pm$ which ensure the Yangian invariance of the graph. We summarize how this is done in the next subsection.

\subsection{Prescription for labels of Lax operators}\label{sec:LaxPrescriptions}

In this section we summarize how one should choose the shifts $\delta^\pm_k$ in the arguments of the monodromy matrix \eq{gre}. The reason why this prescription works is described in  detail for the cross example in section \ref{sec:cross}. The general arguments will be provided in the next section \ref{sec:moves}. For some cases the  prescription was given in \cite{Loebbert:2020hxk} but the situation we consider is more general\footnote{In particular we include the case of multiple internal propagators between two vertices with external legs}.

As a start, for the first external leg with dimension $\Delta$ we choose the labels to be $[\Delta,D/2]$. More generally, the labels of the Lax operator acting on an arbitrary $k$-th external leg will always have the form $[w+\Delta_k,w+D/2]$ for some $w$ (for the very first leg $w=0$). This ensures that indeed the label $\Delta$ of the representation on which the Lax operator acts, as read off from \eq{du}, matches $\Delta_k$.

\begin{figure}[H]
    \hspace{5cm}
\centering
\def\svgwidth{\columnwidth}
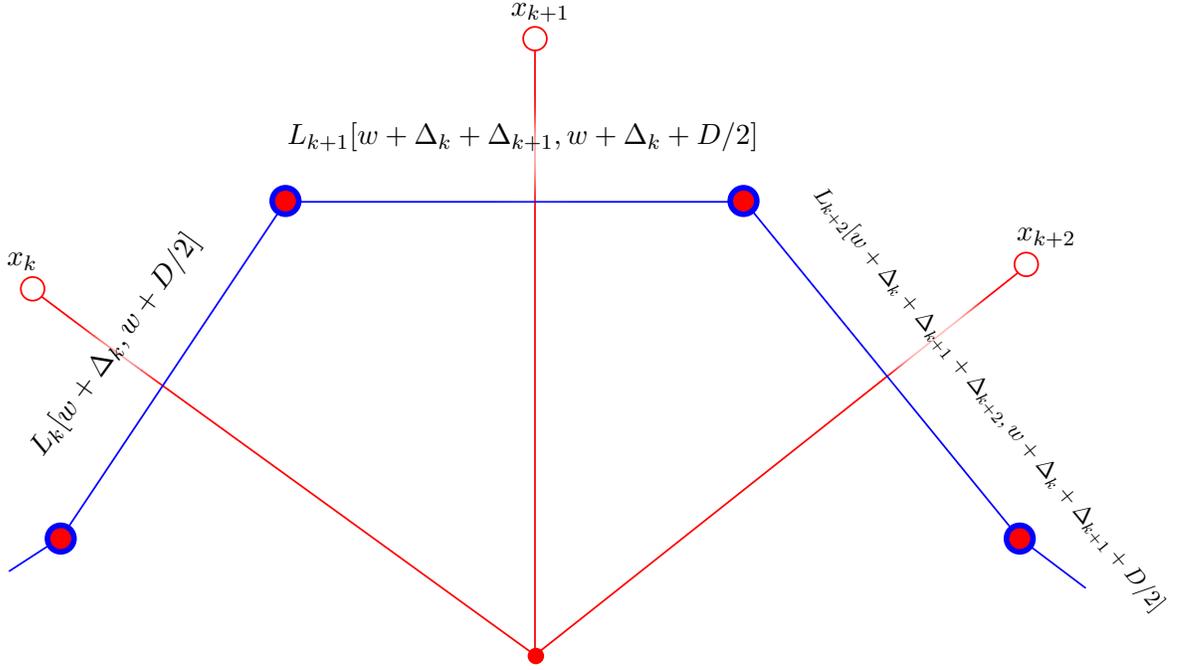
 \caption{Labels for consecutive Lax operators, that share a common vertex. Three of the vertices are depicted with the corresponding Lax operators \eqref{eq:common} }
   \label{fig:prescr1}
\end{figure}

Let us describe how the labels change as we move from leg $k$ to leg $k+1$. First, consider the case when these two legs are attached to the same internal vertex. Denoting by $L_k[w+\Delta_k,w+D/2]$ the Lax operator for the $k$-th leg, we prescribe that the next Lax operator should be chosen as
\beq\label{eq:common}
    L_{k+1}[w+\Delta_k+\Delta_{k+1},w+\Delta_k+D/2] \ .
\eeq
In other words, as we go from leg $k$ to leg $k+1$ we should shift $w\to w+\Delta_k$. We illustrate this prescription on figure \ref{fig:prescr1}. One can see that this reproduces the labels we used for the cross integral in section \ref{sec:cross}. 

\begin{figure}[H]
    \hspace{5cm}
\centering
\def\svgwidth{\columnwidth}
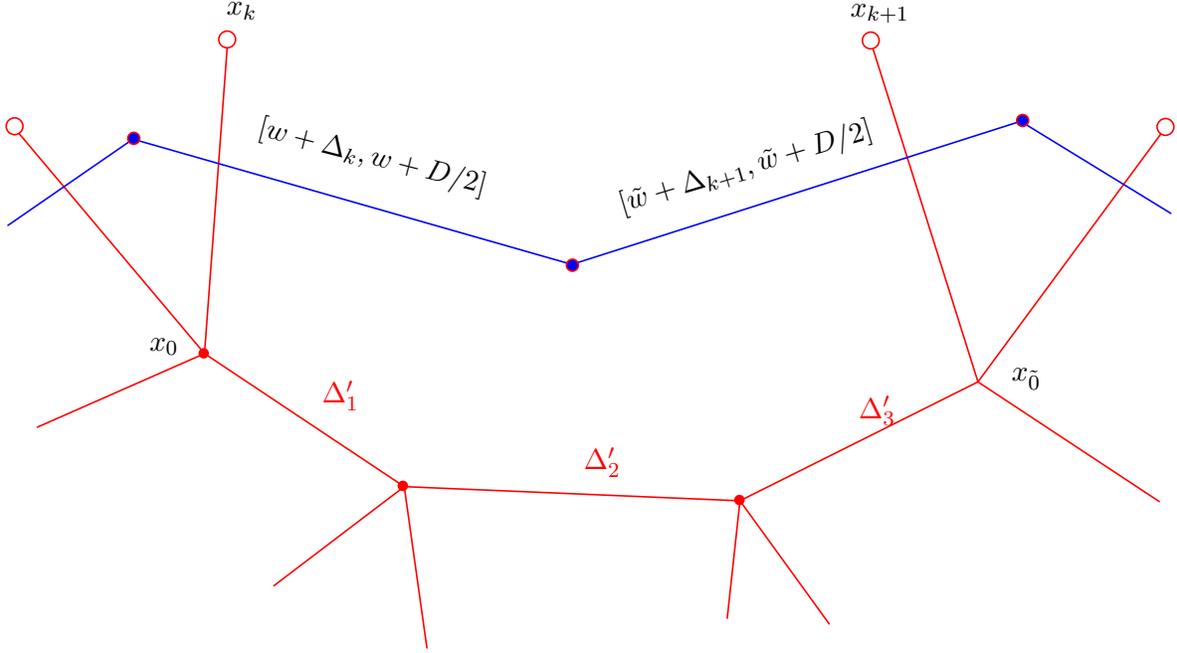
 \caption{Prescription of labels for the Lax operator in the case when consecutive external legs do not share a common internal vertex.  Here $\Tilde{w}$ is given by \eqref{notcommon} with $p=3$.}
   \label{fig:prescr2}
\end{figure}

Alternatively, it could happen that the next leg is attached to a different vertex. Let us label the internal vertex connected to leg $k$ as $0$, and the one connected to leg $k+1$ as $\tilde{0}$. In the most general situation these two vertices $0$ and $\tilde{0}$ may not even be connected directly. However they will always be linked by a chain of propagators between consecutive internal vertices (see figure \ref{fig:prescr2}), starting at vertex  $0$ and ending at vertex $\tilde{0}$. Let us label the dimensions of propagators in this chain as $\Delta_{1}',\dots,\Delta_{p}'$. Then, if the Lax operator for leg $k$ had the form $L_k[w+\Delta_k,w+D/2]$, we find that the one for vertex $k+1$ reads $L_{k+1}[\tilde w+\Delta_{k+1},\tilde w+D/2]$ where
\beq\label{notcommon}
    \tilde w =w+ \Delta_k+\sum_{i=1}^{p}(\Delta_{i}'-D/2) \ .
\eeq
For completeness let us also spell out the corresponding terms in the integral:
\begin{equation}
\begin{split}
      & \ldots L_{k} \left[w+\Delta_k,w+\dfrac{D}{2} \right]   L_{k+1}\left[w+\Delta_k, w+\dfrac{D}{2} \right] \ldots \times
   \\
    & 
     \qquad \times \int \ldots dx_0 \prod_{i'=1}^{p} dx_{i'} dx_{\tilde{0}}  \ldots  \times
    \\
   \qquad & \times \ldots  \dfrac{1}{ \left( x_{0k} \right)^{2\Delta_k}} \dfrac{1}{ \left( x_{01} \right)^{2\Delta'_1}} \left(\prod_{i'=2}^{p-1}\dfrac{1}{\left(x_{i'(i'+1)}\right)^{2\Delta'_{i'}}  }  \right) \dfrac{1}{ \left( x_{p \, \tilde{0} } \right)^{2\Delta_p}} \dfrac{1}{ \left( x_{\tilde{0}(k+1)} \right)^{2\Delta_{k+1}}}\ldots
\end{split}
\end{equation}

Together these rules completely fix the labels one should use to build the monodromy matrix for any graph in the class we consider. To compare this with the $D=4$ fishnet case we notice, that for $\Delta_k=1$ and $D=4$ our prescription turns into the one described in \cite{Chicherin:2017cns,Chicherin:2017frs}. That is easiest to see for simple graphs with convex corners:  when we follow along single external legs the parameters stay constant i.e. $\tilde{w}=w$ in \eqref{notcommon}. On the other hand, when we encounter a vertex with two external legs at a turn we have $[w+1,w+2] \rightarrow [w+2,w+3]$.

\subsection{Moving the lasso}\label{sec:moves}

The claim of Yangian invariance relies on the possibility to reduce the Lax chain that constitutes the lasso to the identity operator. This is achieved via a series of applications of the intertwining relations, pulling all the propagators through the Lax chain. This procedure is represented graphically as pulling the lasso through the diagram, resulting in effectively removing it entirely at the end. The graphical representation in the $D=4$ fishnet case is given in  \cite{Chicherin:2017cns,Chicherin:2017frs}. Here we will generalize the graphical moves to accommodate the diverse graph  geometries.

\begin{figure}[h]
 \centering
   \includegraphics[scale=0.75]{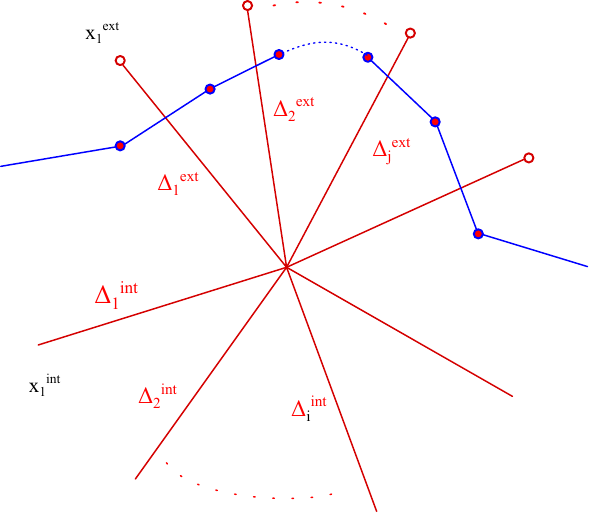}
   \quad
   \includegraphics[scale=0.75]{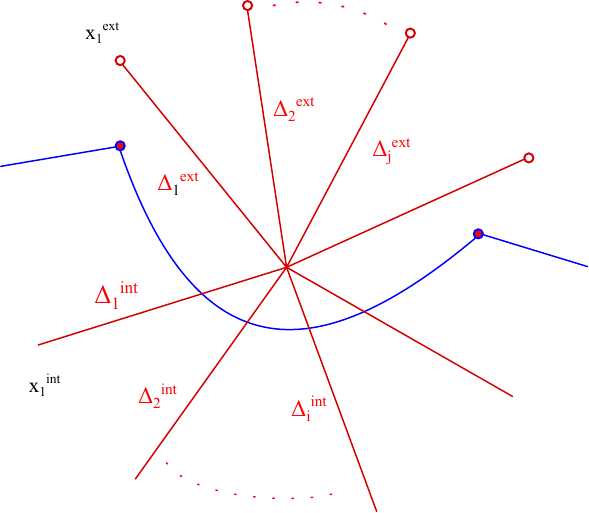} \caption{The first transformation rule that describes flipping the lasso from external legs to internal legs. This should be compered with fig. 4 in \cite{Chicherin:2017frs} and fig. 9 in \cite{Chicherin:2017cns}. }
   \label{fig:move1}
 \end{figure}

Consider a single boundary vertex which is labelled by the coordinate $x_0$. As it lies on the boundary of the diagram, this vertex has neighbouring vertices of two types: external $x^{\mathrm{ext}}_i$ and internal $x^{\mathrm{int}}_j$. The Lax operators initially act on the external vertices with labels prescribed as in section \ref{sec:LaxPrescriptions}. In fact we 
can flip the lasso in such a way that it will now act on the internal vertices instead, as depicted on figure \ref{fig:move1}.

This transformation generalizes the rules on fig. 4 in \cite{Chicherin:2017frs} and fig. 9 in \cite{Chicherin:2017cns}. In particular for the fishnet graph there are two possibilities with either one or two or three external vertices. 
\\\\
Figure \ref{fig:move1} is the graphical representation of the following equality:
\begin{equation}\label{LassoMove1}
\begin{split}
        \int d^D & x_0  \left(\prod_{i=1}^{n_{\mathrm{ext}} }   L_i\left[ w+\sum_{k=1}^{i} \Delta^{\mathrm{ext}}_{k} , w+\dfrac{D}{2} +\sum_{k=1}^{i-1} \Delta^{\mathrm{ext}}_{k}  \right] \right) \cdot 
        \prod_{i=1}^{n_{\mathrm{ext}} } \dfrac{1}{(x^{\mathrm{ext}}_{0i})^{2\Delta^{\mathrm{ext}}_{i}}}
        \cdot \prod_{j=1}^{n_{\mathrm{int}} } \dfrac{1}{(x^{\mathrm{int}}_{0j})^{2\Delta^{\mathrm{int}}_{j}} } =
        \\
        =& \prod_{i=1}^{n_{\mathrm{ext}}-1} \left[ w+\dfrac{D}{2} +\sum_{j=1}^{i}\Delta^{\mathrm{ext}}_j\right]
        \times 
        \\
        \times \ \ &
        \int d^D x_0  \prod_{i=1}^{n_{\mathrm{ext}} } \dfrac{1}{(x^{\mathrm{ext}}_{0i})^{2\Delta^{\mathrm{ext}}_{i}}}  \cdot L_{0}\left[ w+\dfrac{D}{2} , w+ \sum_{j=1}^{n_\mathrm{ext}} \Delta^{\mathrm{ext}}_{j}  \right] 
      \cdot \prod_{j=1}^{n_{\mathrm{int}} } \dfrac{1}{(x^{\mathrm{int}}_{0j})^{2\Delta^{\mathrm{int}}_{j}} } \ .
\end{split}
\end{equation}
Here $n_\mathrm{ext}$ and $n_\mathrm{int}$ are the number of external and internal vertices respectively. Thus, the prescription for the parameters of the Lax operator in sec. \ref{sec:LaxPrescriptions} for legs attached to the same vertex is chosen in such a way that formula \eqref{LassoMove1} works. 

\begin{figure}[H]
 \centering
   \includegraphics[scale=0.75]{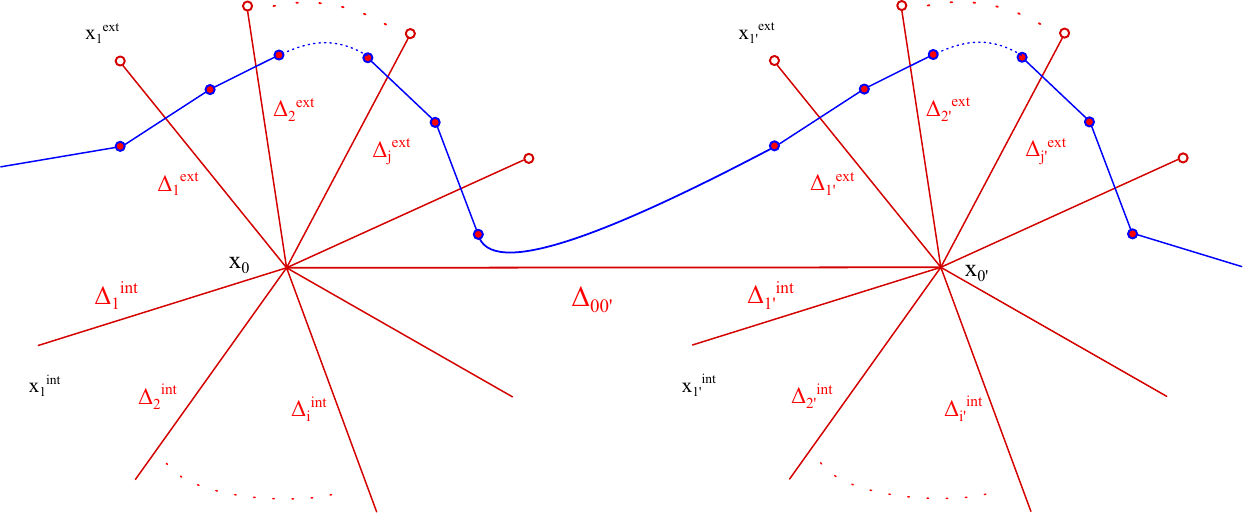}
   \caption{The Lax operator chain for two neighbouring internal vertices $x_0$ and $x_{0'}$. For simplicity we consider the $p=1$ case of \eqref{notcommon}.}
      \label{fig:nextmove1}
 \end{figure}

On the other hand it is important that the labels between two consecutive sets of external lines are in agreement, which enables the second prescription in section \ref{sec:LaxPrescriptions}. Indeed, consider now two vertices $x_0$ and $x_0'$ with external vertices $x_k^{\mathrm{ext}}$ and $x_{k'}^{\mathrm{ext}}$ and internal vertices  $x_j^{\mathrm{int}}$ and $x_{j'}^{\mathrm{int}}$, drawn on figure \ref{fig:nextmove1}. 

\begin{figure}[H]
 \centering
   \includegraphics[scale=0.75]{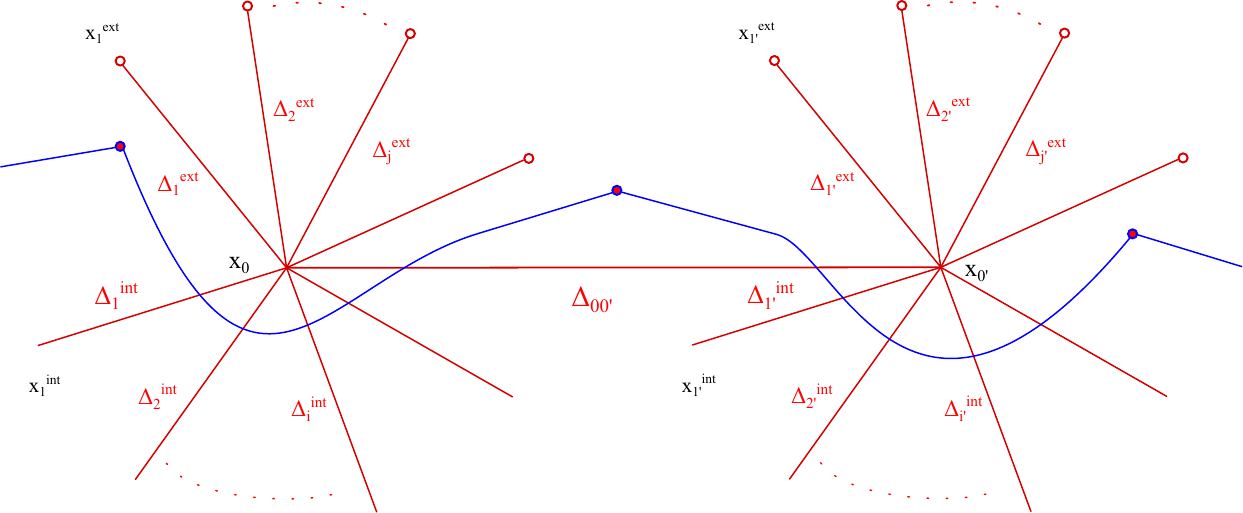}
   \caption{The Lax chain after transformations of fig. \ref{fig:move1} was applied to each vertex $x_0$ and $x_0'$. The figure corresponds to the l.h.s of \eqref{eq:00'intertwiner}} 
      \label{fig:nextmove2}
 \end{figure}

\begin{figure}[h]
 \centering
   \includegraphics[scale=0.75]{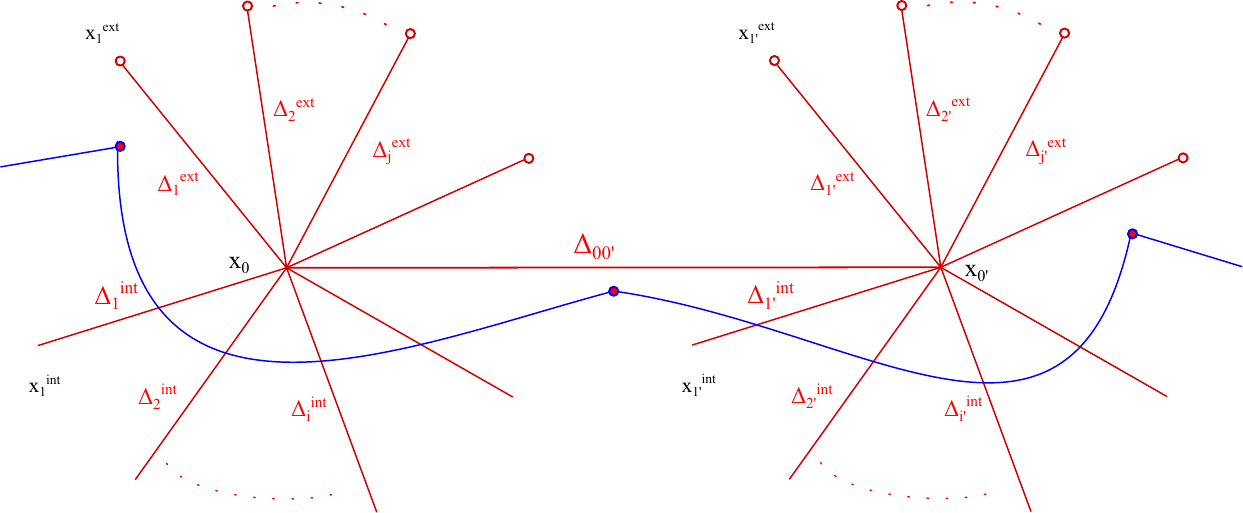}
   \caption{The Lax operator has been pulled of the $00'$ edge, as on the r.h.s of \eqref{eq:00'intertwiner}. To be able to make this transformation we should impose \eqref{notcommon}}
   \label{fig:nextmove3}
 \end{figure}

The Lax parameters for each vertex are as prescribed but with parameters $w$ and $\tilde{w}$ as in section \ref{sec:LaxPrescriptions}. Suppose we make the transformation for each vertex \eq{LassoMove1}, which will leave us with the configuration on figure \ref{fig:nextmove2}.
 One of  the vertices internal for $x_0$ is $x_0'$ itself and vice versa. Let us look at the propagator connecting those two vertices $x_{00'}$ which has dimension $\Delta_{00'}$ and consider the intertwining relation:
\begin{equation}\label{eq:00'intertwiner}
\begin{split}
        L_{0}&\left[  w+\dfrac{D}{2} , w+ \sum_{j=1}^{n_\mathrm{ext}} \Delta^{\mathrm{ext}}_{j}  \right] \cdot L_{0'}\left[ \tilde{w}+\dfrac{D}{2} , \tilde{w}+ \sum_{j'=1}^{n'_\mathrm{ext}} \Delta^{\mathrm{ext}}_{j'}  \right] \dfrac{1}{\left(x_{0 0'}\right)^{2\Delta_{00'}}} = 
        \\
        &=  \dfrac{1}{\left(x_{0 0'}\right)^{2\Delta_{00'}}} L_{0}\left[ \tilde{w}+ \sum_{j'=1}^{n'_\mathrm{ext}} \Delta^{\mathrm{ext}}_{j'} , w+ \sum_{j=1}^{n_\mathrm{ext}} \Delta^{\mathrm{ext}}_{j}  \right] \cdot L_{0'}\left[ \tilde{w}+\dfrac{D}{2} ,  w+\dfrac{D}{2}  \right] \ .
\end{split}
\end{equation}
In order for this relation to hold we have to set:
\begin{equation}
    w+\dfrac{D}{2} = \tilde{w}+ \sum_{j'=1}^{n'_\mathrm{ext}} \Delta^{\mathrm{ext}}_{j'}  + \Delta_{00'}
\end{equation}
which is nothing but the case $p=1$ of the prescription \eqref{notcommon} announced above. By using this relation we will finally have the configuration on figure \ref{fig:nextmove3}, which means we have pulled the lasso through the $00'$ edge.
The fact that the graphical moves actually result in the lasso being completely removed is rather nontrivial. It relies on the prescription of Lax labels being consistent with the dimensions of the external and internal vertices. We prefer to think that there are two types of relations between dimensions that appear. Relations of the first type, which we call local, represent only the fact that the diagram is conformal, i.e. the sum of conformal dimensions adds up to $D$. The second type of relations are in contrast referred to as non-local and are a direct consequence of the Loom construction: since the initial lattice is made up from straight lines there are geometric relations between far separated angles and hence between conformal dimensions. Examples of such relations are discussed in sections \ref{sec:comments} and \ref{sec:example6legs}.

\begin{figure}[h]
 \centering
   \includegraphics[scale=0.75]{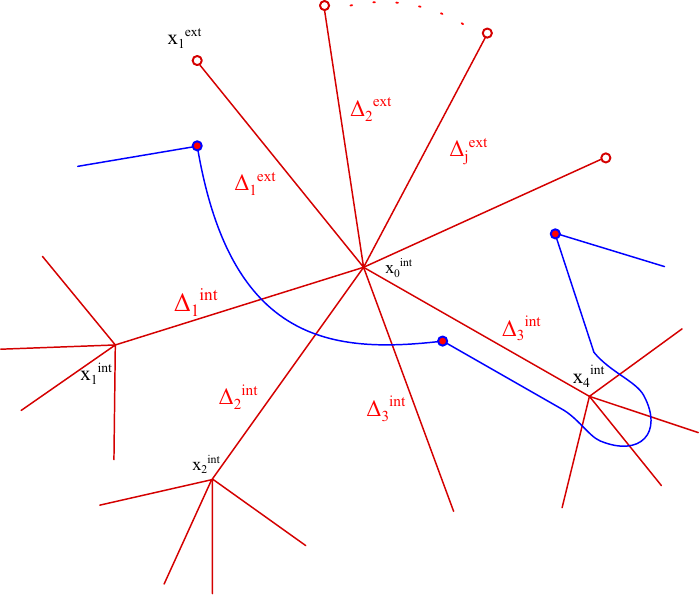}
   \caption{The  Lax chain of the lasso after transformations of fig. \ref{fig:move1} (formula \eqref{LassoMove1}) has been applied to $x_0$ and $x_{n_{\mathrm{int}}}^{\mathrm{int}}$, for $n_{\mathrm{int}}=4$. This figure demonstrates \eqref{ultimatemove}. As clearly seen from the picture, after applying it to all vertices $x_j^{\mathrm{int}}$ the lasso won't act on the coordinate $x_0$ anymore. }
   \label{fig:ultimatemove}
 \end{figure}

Let us now prove that the lasso can be moved past a single vertex completely. To do this, consider the same vertex $x_0$. For each line going to an ''internal''  vertex consider now its neighbours. For a vertex $x^{\mathrm{int}}_j$ we label them by the coordinates $y_{i,j}$. Assume for simplicity that the graph is big enough, such that the lasso does not pass through any of the mentioned edges other than $x_{0i}$. The setup is represented on figure \ref{fig:ultimatemove}. Now first pull the lasso through the $x_0$ vertex using \eqref{LassoMove1}.  We still have the $L_0$ Lax operator on the r.h.s. which we want to get rid of, such that the $x_0$ coordinate is completely out of the game. To do this first use the conformal nature of the diagram and rewrite the parameters of the Lax operator $L_0$ in terms of dimensions of internal legs:
\begin{equation}
    L_{0}\left[ w+\dfrac{D}{2} , w+ \sum_{j=1}^{n_\mathrm{ext}} \Delta^{\mathrm{ext}}_{j}  \right] = L_0\left[w+\dfrac{D}{2},w+D-\sum_{j=1}^{n_{\mathrm{int} }} \Delta^{\mathrm{int} }_j \right] \, .
\end{equation}
Then use the move on fig. \ref{fig:move1} again but for the last (with label $n_{\mathrm{int}}$)internal vertex  $x_{n_{\mathrm{int}}}^{\mathrm{int}}$:
\begin{equation}\label{ultimatemove}
\begin{split}
 & \int d^D x_0  d^D x_{n_{\mathrm{int}} }^{\mathrm{int}}  L_0\left[w+\dfrac{D}{2},w+D-\sum_{j=1}^{n_{\mathrm{int}}} \Delta^{\mathrm{int} }_j \right] \dfrac{1}{(x_{0n_{\mathrm{int}}}^{\mathrm{int}})^{2\Delta_{n_{\mathrm{int}}}^{\mathrm{int}}} } \cdot  \prod_{j= 1}^{n_{\mathrm{int}}-1 } \dfrac{1}{(x^{\mathrm{int}}_{0j})^{2\Delta^{\mathrm{int}}_{j}} }  \times
 \\
 &\times \prod_{i=1}^{m} \dfrac{1}{ \left( 
 x^{\mathrm{int}}_{n_{\mathrm{int}}} - y_{i,n_{\mathrm{int}}} \right)^{2\Delta_{i} } } \cdot \ldots=
 \\[1pt]
 &=\int d^D x_0 d^D x_{n_{\mathrm{int}}}^{\mathrm{int}} 
  \dfrac{1}{(x_{0n_{\mathrm{int}}}^{\mathrm{int}})^{2\Delta_{n_{\mathrm{int}}}^{\mathrm{int}}} } \dfrac{1}{\left[\tilde{w}+\dfrac{D}{2}\right]} L_{n_{\mathrm{int}}}\left[\tilde{w}+\dfrac{D}{2}-\Delta^{\mathrm{int}}_{n_{\mathrm{int}}}, \tilde{w} \right] \times
 \\[1pt]
 &\times  L_0\left[w+\dfrac{D}{2} , w+D-\sum_{j=1}^{n_{\mathrm{int}}} \Delta^{\mathrm{int} }_j +  {\color{blue} \Delta^{\mathrm{int} }_{n_{\mathrm{int}} }}  \right]  
 \prod_{j =1}^{n_{\mathrm{int}}-1 } \dfrac{1}{(x^{\mathrm{int}}_{0j})^{2\Delta^{\mathrm{int}}_{j}} } \cdot 
 \prod_{i=1}^{m} \dfrac{1}{ \left( 
 x^{\mathrm{int}}_{n_{\mathrm{int}}} - y_{i,n_{\mathrm{int}}} \right)^{2\Delta_{j} } } \cdot \ldots
 \end{split}
\end{equation}

This transformation is represented on figure \ref{fig:ultimatemove}. The parameter $\tilde{w}$ is expressed through $w$ as:
\begin{equation}
    \tilde{w}=w+\dfrac{D}{2}-\sum_{j=1}^{n_{\mathrm{int}}-1} \Delta_j \ .
\end{equation}
The key observation is the shift of parameters in the $L_0$ Lax operator highlighted in {\color{blue} blue} in \eqref{ultimatemove}. After the transformation we are left with the same Lax operator $L_0$ with the parameters now involving one less internal vertex:
\begin{equation}
    w+D-\sum_{j=1}^{n_{\mathrm{int}}} \Delta^{\mathrm{int} }_j +  {\color{blue} \Delta^{\mathrm{int} }_{n_{\mathrm{int}} }} = w+ D- \sum_{j=1}^{n_{\mathrm{int}}-1} \Delta^{\mathrm{int} }_j \ .
\end{equation}It is clear that the same shift will happen for all internal vertices, if we do the same transformation vertex by vertex. Hence after going through all internal vertices $x_j^{\mathrm{int}}$ the resulting Lax operator acts on identity, since all the $x_{0j}$ propagators have been pulled through, with the parameters consistently shifted:
\begin{equation}
   L_0\left[w+\dfrac{D}{2} , w+D-\sum_{k} \Delta^{\mathrm{int} }_k +  {\color{blue} \sum_{j} \Delta^{\mathrm{int} }_j } \right] \cdot 1  =  L_0\left[w+\dfrac{D}{2},w +D \right] \cdot 1 = \left[w+D\right] \cdot 1 \ .
\end{equation}
For all the internal vertices $x_j^{\mathrm{int}}$ we now are at the previous step, where we have a single Lax operator action on $x_j^{\mathrm{int}}$. Notice, that the Lax operators the appear for the internal vertices are with appropriate parameters. For example, for the last vertex as in \eqref{ultimatemove}, the Lax operator can be rewritten as:
\begin{equation}
    L_{n_{\mathrm{int}}}\left[\tilde{w}+\dfrac{D}{2}-\Delta^{\mathrm{int}}_{n_{\mathrm{int}}}, \tilde{w} \right]= L_{n_{\mathrm{int}}}\left[(\tilde{w}-\Delta^{\mathrm{int}}_{n_{\mathrm{int}}})+\dfrac{D}{2}, (\tilde{w}-\Delta^{\mathrm{int}}_{n_{\mathrm{int}}}) + \Delta^{\mathrm{int}}_{n_{\mathrm{int}}}\right] \ .
\end{equation}
Since the corresponding propagator is now \emph{external} w.r.t to vertex $ x_{n_{\mathrm{int}}}$ we are exactly in the situation as in the r.h.s of formula \eqref{LassoMove1}. 
We then repeat the same transformations to consistently remove the Lax chain from each vertex. 

The fact that this sequence of moves indeed leads to the lasso being completely pulled off the diagram is a more involved condition on the conformal dimensions.  This is because the process of pulling of a single vertex or even between two adjacent vertices requires only the conformal condition and peculiar relations between Lax parameters. These other ones should be the conditions that make the moves on different vertices in distant parts of the diagram consistent with each other. The origin of these relations seems to be the fact that the diagram is drawn on the dual space to the loom --  a lattice made up from straight lines.
%--------------%

%--------------%

%--------------%

%--------------%

%-------------- %

%--------------%

As a final remark, we expect that the steps we have proven above should also be valid for the generalised case when internal vertices are placed inside open faces as described in section \ref{sec:gen} and we have checked this for a number of examples. This is also natural as likely the introduction of these vertices may be viewed as a kind of analytic continuation in the space of parameters (i.e. from the case when the face is closed to the case when it is open). We leave details of the proof for this case to the future.

\subsection{Example of applying the lasso moves}
To illustrate the techniques described above we provide an example of a specific graph which has various non-trivial features that were present in the general construction. However, since writing out the whole Lax chain with all the arguments is still too lengthy we will shorten our presentation by only spelling out those terms that appear in the relevant transformations and use shorthand notation for the parameters of the Lax operators.
\begin{figure}[H]
    \centering
    \includegraphics[width=\linewidth]{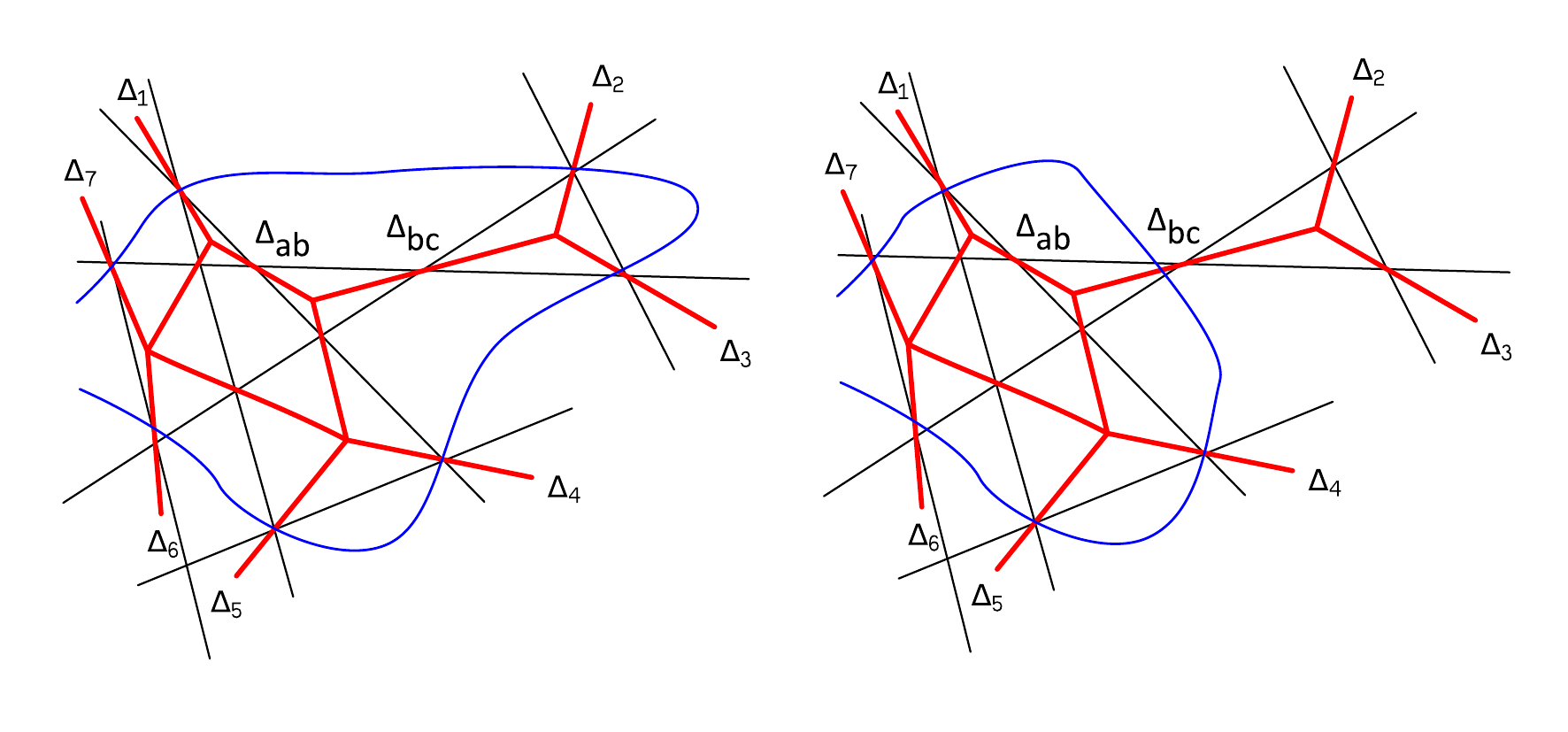}
    \caption{Example of a nontrivial lasso move.}
    \label{fig:loomexample1}
\end{figure}
The graph and its Lax operator chain is shown on the left in fig. \ref{fig:loomexample1}. We denote by $\Delta_i$ the conformal dimensions of the external legs, and label their coordinates by $x_i$. We also label the coordinates of 3 of the internal vertices by $x_a,x_b,x_c$ according to the figure, and the corresponding dimensions of propagators between them as $\Delta_{ab},\Delta_{bc}$. Vertex $c$ is connected to $x_2$ and $x_3$, while vertex $a$ is connected to $x_1$.
\\

In what follows we will illustrate what happens to parts of the Lax operator chain which act on these vertices. The relevant part of the Lax chain and the integrand are:
\begin{equation}
    \ldots \cdot L_3[\delta_3^+,\delta_3^-] L_2[\delta_2^+,\delta_2^-]L_1[\delta_1^+,\delta_1^-] \ldots \dfrac{1}{x_{a1}^{2\Delta_1}x_{ab}^{2\Delta_{ab}} x_{bc}^{2\Delta_{bc}} x_{c2}^{2\Delta_2} x_{c3}^{2\Delta_3}}  \ldots
\end{equation}
Here the shifts of the spectral parameter are calculated according to the prescriptions in sec. \ref{sec:LaxPrescriptions}:
\begin{equation}
\begin{split}
        &(\delta^+_1 ,\delta_1^-) = (\Delta_1, D/2), \  \quad (\delta^+_2 ,\delta_2^-) = (\Delta_1+\Delta_2+\Delta_{ab}+\Delta_{bc}-D, \Delta_1+\Delta_{ab}+\Delta_{bc}-D/2)
        \\
        &(\delta^+_3 ,\delta_3^-) =(\delta_2^++\Delta_3,\delta_2^-+\Delta_2)
\end{split}
\end{equation}

\begin{figure}[H]
    \centering
    \includegraphics[width=\linewidth]{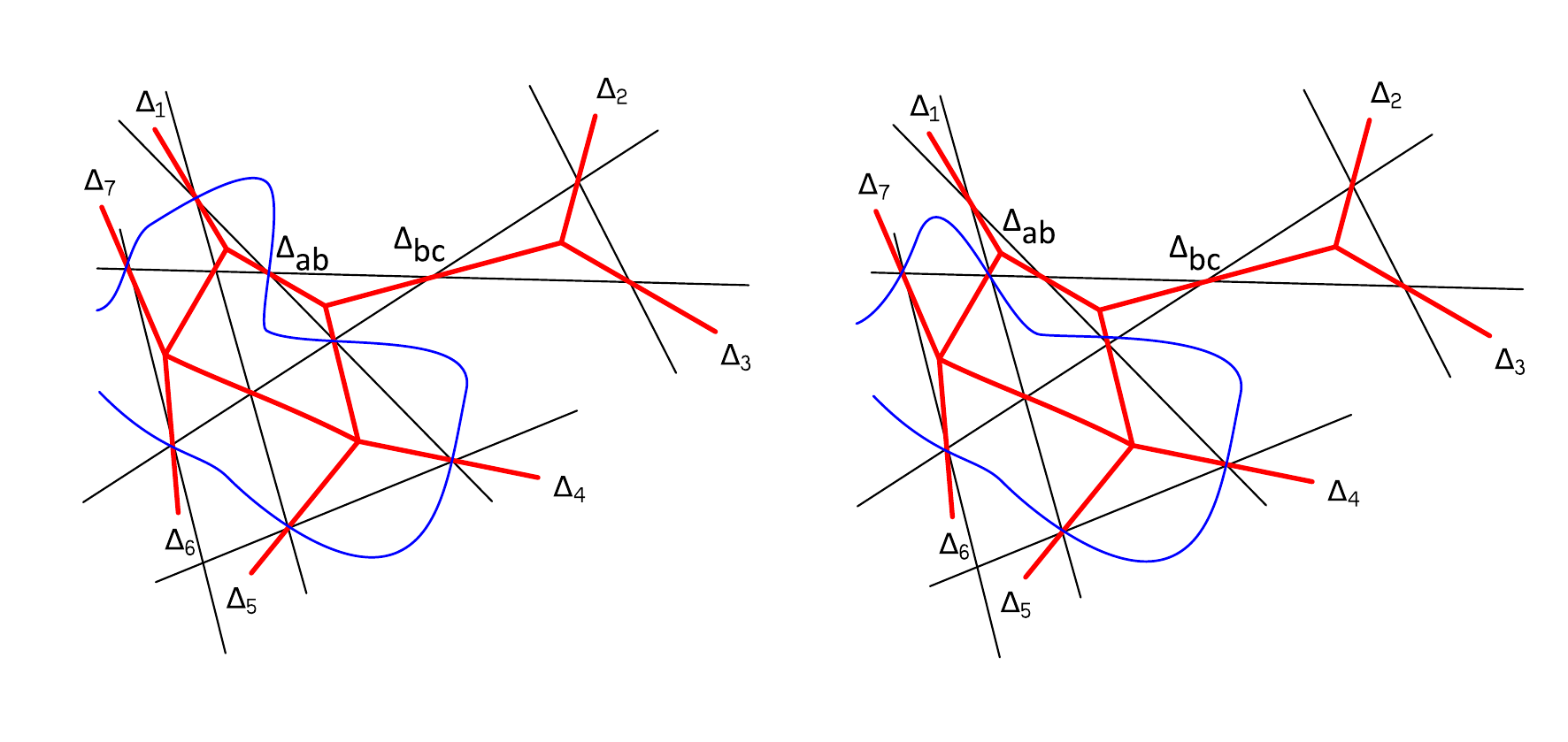}
    \caption{Example of a nontrivial lasso move -- continued from figure \ref{fig:loomexample1}.}
    \label{fig:loomexample 2}
\end{figure}
To pull the Lax chain through the diagram we go through a sequence of transformations:
\begin{itemize}
    \item First apply the flipping transformation \eqref{LassoMove1} of fig. \ref{fig:move1} to the legs $x_2,x_3$ of vertex $x_c$. This is illustrated on the right in fig. \ref{fig:loomexample1}. According to \eqref{LassoMove1} we get (ignoring the factors):
    \begin{equation}
        \begin{gathered}
                        \ldots L_3[\delta^+_3,\delta_3^-] L_2[\delta^+_2,\delta_2^-]  \ldots \dfrac{1}{x_{2c}^{2\Delta_2}x_{3c}^{2\Delta_3}} \ldots\\
            \downarrow
            \\
            \ldots  L_c[\delta_2^+-\Delta_1+D/2,\delta_2^--D/2+\Delta_2+\Delta_3]  \ldots  \dfrac{1}{x_{bc}^{2\Delta_{bc}}} \ldots
        \end{gathered}
    \end{equation}
    Note that after this transformation the Lax operator still acts on vertex $x_c$. 
    \item Next we apply the flipping transformation \eqref{LassoMove1}
     to the vertex $x_b$, treating the propagator $bc$ as an external leg. The result is drawn in fig. \eqref{fig:loomexample 2} on the left. This is nothing but the transformation described in \eqref{ultimatemove} and fig. \ref{fig:ultimatemove}. We get:
   \begin{equation}
        \begin{gathered}
         \ldots L_c[\delta_c^+,\delta^-_c]  \ldots  \dfrac{1}{x_{bc}^{2\Delta_{bc}}} \ldots
         \\
         \downarrow
         \\
         \ldots L_b[\delta_c^+-\Delta_{bc}+D/2,\delta_c^--D/2+\Delta_{bc}]  \ldots  \dfrac{1}{x_{ab}^{2\Delta_{ab}}} \ldots
        \end{gathered}
    \end{equation}
    where we denoted
    \begin{equation}
        (\delta_c^+,\delta^-_c)=(\delta_2^+-\Delta_2+D/2,\delta_2^--D/2+\Delta_2+\Delta_3)
    \end{equation}
    With this step we pulled the lasso completely through the vertex $x_c$. 
    \item On the next step we apply the flipping move to vertex $a$ with the external leg stretching to vertex $x_1$ and obtain as a result
    \begin{equation}
       \ldots  L_a[D/2,\Delta_1] \ldots  \dfrac{1}{x_{ab}^{2\Delta_{ab}}} \ldots
    \end{equation}
    \item The next step demonstrates a non-trivial consistency condition that appears as a result of the Loom construction. Notice that now both $L_b$ and $L_a$ act on the propagator $\frac{1}{x_{ab}^{2\Delta_{ab}}}$. Moreover, $L_b$ was obtained after we pulled the lasso through the vertex $x_c$. Hence it contains traces of these transformations in the shift parameters, namely, the dimensions $\Delta_{bc},\Delta_1,\Delta_2$. However, the consistency conditions from the Loom guarantee the parameters are now set in such a way that we can use the intertwining relation. Namely, consider the expression:
    \begin{equation}
        \ldots   L_b[\delta_c^+-\Delta_{bc}+D/2,\delta_c^--D/2+\Delta_{bc}] L_a[D/2,\Delta_1] \ldots  \dfrac{1}{x_{ab}^{2\Delta_{ab}}} \ldots
    \end{equation}
    and notice that the relation between the first shift of the $L_b$ operator and the second shift of the $L_a$ operator is:
    \begin{equation}
    \begin{split}
           &(\delta_c^+-\Delta_{bc}+D/2) - \Delta_1 = \delta_2^{+}-\Delta_2+D/2-\Delta_{bc}+D/2-\Delta_1
=
\\&= \Delta_{ab}
    \end{split}
    \end{equation}
    which is precisely the relation required to apply the intertwining identity for the propagator $\dfrac{1}{x_{ab}^{2\Delta_{ab}}}$.
    Doing this we obtain the right configuration in fig. \ref{fig:loomexample 2}.
\end{itemize}
Continuing the series of transformations in a similar fashion we pull the lasso through the whole diagram and obtain the eigenvalue equation for the Lax chain.

\subsection{Cyclicity and computation of the eigenvalue}

Using the lasso procedure described above, one can compute the eigenvalue of the monodromy matrix directly for any particular graph. Alternatively, one can in fact write down a difference equation which fixes the eigenvalue in terms of the labels $\delta_n^\pm$ of the Lax operators. This can be done following \cite{Chicherin:2017cns} by making use of the properties \eq{inv1}, \eq{inv2} which ultimately allow one to cyclically reorder the Lax operators inside the monodromy matrix while at the same time shifting their labels $\delta_n^\pm$ by some constants. The derivation is an immediate generalisation of the discussion from appendix A of \cite{Chicherin:2017cns} (with minimal modifications due to the $D$-dependent shifts in \eq{inv1}, \eq{inv2}). As a result, we find that the eigenvalue $\lambda(u)$ satisfies the relation
\beq
\label{ld}
    \frac{\lambda(u)}{\lambda(u-D)}=\frac{P(u)}{P(u-D/2)} \ ,
\eeq
where $P$ is a polynomial encoding the values of $\delta_n^\pm$,
\beq
    P(u)=\prod_{j=1}^n(u+\delta_j^+)(u+\delta_j^-) \ .
\eeq
This equation completely fixes the eigenvalue $\lambda$. In particular, one can use it to find its large $u$ expansion which will be useful in section \ref{sec:diff}. We find
\beq
    \lambda(u) = u^n+\frac{1}{2}u^{n-1}\sum_{k=1}^n\hat\delta_k
    +\frac{1}{4}u^{n-2}\left[\sum_{i<j}\hat\delta_i\hat\delta_j-\frac{1}{2}\sum_i\hat\Delta_i\right]+\dots
\eeq
where we denoted (following \cite{Chicherin:2017cns}) 
\beq
    \hat \delta_k=\delta_k^++\delta_k^-+D/2 \ , \ \ \hat\Delta_i=\Delta_i(\Delta_i-D) \ .
\eeq
We recall that $\Delta_k=\delta_k^+-\delta_k^-+D/2$.

\section{Differential equations from Yangian symmetry and examples}

\label{sec:diff}

In this section we derive differential equations following from Yangian symmetry and demonstrate them for several examples.

In general the Yangian of a Lie algebra (in our case, the Lie algebra of the conformal group in $D$ dimensions) can be described in terms of level-0 generators $J^A$ and level-1 generators ${\widehat J}^A$, see e.g. \cite{Chicherin:2017cns} for a more detailed discussion and \cite{Chicherin:2022nqq} for a review of how the monodromy matrix encodes these generators in its large $u$ expansion. This is known as the first realization of the Yangian. The level-0 generators are those of the original Lie algebra and satisfy the commutation relations
\beq
    [J^A,J^B]=f^{AB}_{\ \ \ \  C}J^C
\eeq
where $f^{AB}_{\ \ \ \  C}$ are the structure constants. In our case these generators are given by a sum of individual operators acting on each external leg of the graph,
\beq
    J^A=\sum_{k=1}^n J^A_k \ .
\eeq
The level-1 generators are bilocal and can be written as
\beq
\label{jh}
    \hat J^A=\frac{1}{2}f^A_{\ \ \ BC}\sum_{j<k}J^C_jJ^B_k+\sum_kv_kJ^A_k
\eeq
in terms of some evaluation parameters $v_k$. All other generators of the infinite-dimensional Yangian algebra can be obtained as some polynomial combinations of $J^A$ and ${\widehat J}^A$.

In our case the explicit form of the Yangian generators that annihilate our graph, together with the evaluation parameters $v_k$, can be read off from the first few orders of the large $u$ expansion of the monodromy matrix \eq{gre} and the eigenvalue $\lambda(u)$. The latter can be obtained from \eq{ld} while the former can be found by using the explicit form of the Lax matrix \eq{lax}. A detailed derivation is given in section 8.3 of  \cite{Chicherin:2017cns}. Extending it to our case with generic dimension $D$, we find that the evaluation parameters are given in terms of the shifts in Lax operators in \eq{gre} as
\beq
\label{vres}
    v_k=\frac{1}{2}\sum_{j\neq k}(\delta_j^+ + \delta_j^-+D/2) \ .
\eeq

The conformal generators $J^A$ and the level-1 Yangian generators ${\widehat J}^A$ both annihilate our Feynman integral. While the first statement amounts to conformal invariance of the graph, the second one gives further nontrivial constraints. As discussed in \cite{Chicherin:2017cns} it is sufficient to consider only the case of the momentum generators, i.e. ${\widehat J}^A={\widehat P}^\mu$ (the other generators do not give new independent equations). Explicitly, this generator reads \cite{Chicherin:2017cns}
\beq
\label{ph}
    {\widehat P}^\mu=-\frac{i}{2}\sum_{j<k}[(L_j^{\mu\nu}+g^{\mu\nu}D_j)P_{k,\nu}-(j\leftrightarrow k)]+\sum_j v_jP_j^\mu
\eeq
where the most nontrivial graph-dependent part is the last term. Plugging in the conformal generators from \eq{conf1}, \eq{conf2} one can write it as an explicit differential operator that annihilates our graph,
\beq
    {\widehat P}^\mu\ket{\rm graph}=0 \ .
\eeq
Let us mention that in \eq{jh} we can shift all $v_k$ by the same value $v_k\to v_k+c$ since the level-0 generators $J^A$ annihilate our state. For example, by doing this we can always set $v_1= 0$.

Although the derivation of the differential equations following from Yangian symmetry relies on the explicit form of the Lax matrix \eq{lax} which we discussed only for even dimension $D$, the only dependence on $D$ in the end is parametric, contained in the labels $\delta_k^\pm$ and various shifts proportional to $D$ such as in \eq{vres}. This suggests that the resulting differential equations in fact hold in any dimension, though it would be important to establish this more rigorously.

Since the graphs are conformal, they must evaluate to some nontrivial functions of the conformal cross-ratios times explicit spacetime dependent prefactors. Plugging this representation into the differential equations, one can separate the terms with different coordinate dependence and obtain differential equations on the independent components written now only in terms of the cross ratios. This rewriting applies to our general case as well and we refer to \cite{Loebbert:2022nfu}  for more details.

As a result, we see that in order to write the differential equations for the graph we simply read off the labels $\delta_k^\pm$ and plug them into \eq{vres}, \eq{ph}. Below we illustrate this on an example.

\subsection{Example: square with 6 legs}\label{sec:example6legs}

 \begin{figure}[h]
 \centering
   \includegraphics[scale=0.9]{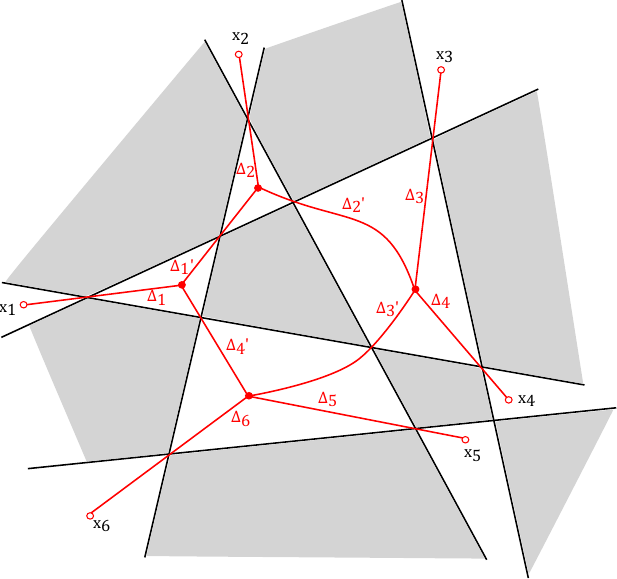}  \caption{Square with 6 legs. }
   \label{fig:sq6}
 \end{figure}

As a first example, let us consider a graph that looks like a square with 6 external legs (figure \ref{fig:sq6}). It is parameterised by 10 scaling dimensions, labelled as $\Delta_1,\dots,\Delta_6$ and $\Delta_1',\dots,\Delta_4'$ on the figure. Imposing that the dimensions at each vertex sum up to  $D$ gives four independent relations. In addition to them we have a further 'nonlocal' relation
\beq
\label{sd3}
    \Delta_6+\Delta_2+\Delta_5=D 
\eeq
which one can deduce from the geometry and the rule \eq{da} linking dimensions to angles. One can show that this exhausts all independent relations between the dimensions coming from the loom construction in this case\footnote{One way to do it is to parameterise each line in terms of its angle w.r.t. the horizontal direction (these angles are clearly completely independent variables) and then express the dimensions through combinations of these angles. }. Notice that from these relations it also follows that we have a relation similar to \eq{sd3},
\beq
     \Delta_4+\Delta_1+\Delta_3=D \ ,
\eeq
as well as
\beq
    \sum_{i=1}^4\Delta_i'=D \ .
\eeq
As a result, we can express all 10 dimensions in terms of 5 parameters, for example in terms of  $\Delta_1,\Delta_2,\Delta_3,\Delta_5$ and $\Delta_1'$ which gives
\beq
\label{sdrel1}
    \Delta_6=D-\Delta_2-\Delta_5 \ , \ \ \ \Delta_4=D-\Delta_1-\Delta_3 \ ,
\eeq
\beq
\label{sdrel2}
     \Delta_2'=D-\Delta_1'-\Delta_2 \ , \ 
    \Delta_3'=-D+\Delta_1+\Delta_1'+\Delta_2 \ , \ \Delta_4'=D-\Delta_1-\Delta_1' \ .
\eeq

Using the results of section \ref{sec:yang}, we can write explicitly the labels for the monodromy matrix of which the graph is an eigenstate. It has the form 
\beqa
\label{sqmon}
&& 
     L_6[\Delta_{(11')}+D/2,\Delta_{(121'5)}]L_5[\Delta_{(121'5)}-D/2,
     \Delta_{(121')}]L_4[D,\Delta_{(13)}+D/2] \times
     \\ \nn && L_3[\Delta_{(13)},\Delta_{1}+D/2]L_2[\Delta_{(121')}-D/2,\Delta_{(11')}]L_1[\Delta_1,D/2]
\eeqa
where we used the notation
\beq
    \Delta_{(a_1 a_2\dots a_p)}=\Delta_{a_1}+\Delta_{a_2}+\dots+\Delta_{a_p} \ .
\eeq
The corresponding evaluation parameters entering the differential equation that follows from Yangian symmetry are obtained from \eq{vres} (we used the freedom of shifting them all by the same constant to set $v_1=0$),
\beqa
&& 
    v_k=\left\{0,-\Delta _1'-\frac{\Delta _1}{2}-\frac{\Delta _2}{2}+D/2,-\frac{\Delta _1}{2}-\frac{\Delta _3}{2}, -\frac{\Delta _3}{2}-D/2,\right.
    \\ \nn && \left. \ \ \ \ \ \ \ \ 
    -\Delta
   _1'-\frac{\Delta _1}{2}-\Delta _2-\frac{\Delta _5}{2}+D/2,-\Delta _1'-\frac{\Delta _1}{2}-\frac{\Delta _2}{2}-\frac{\Delta
   _5}{2}\right\} \ .
\eeqa

\subsubsection{Reduction to 4 legs}

 \begin{figure}[h]
 \centering
   \includegraphics[scale=0.9]{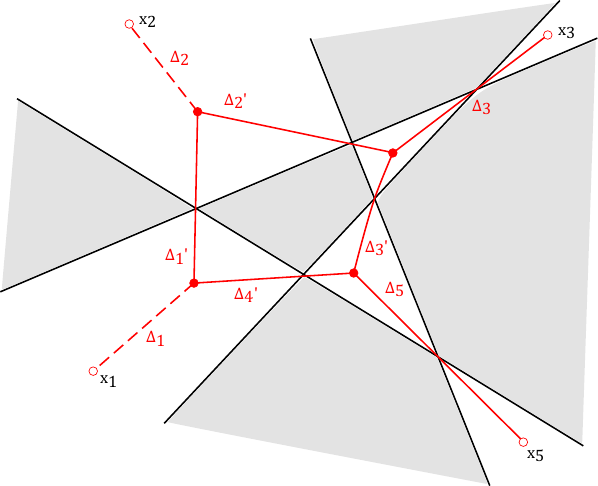}  \caption{Square with 4 legs. }
   \label{fig:sqlab}
 \end{figure}

An interesting limit for this graph with 6 legs is when we set to zero the dimensions $\Delta_6$ and $\Delta_4$, leaving a graph with only one leg coming out of each vertex. At the level of the relations between scaling dimensions \eq{sdrel1}, \eq{sdrel2} and of the monodromy matrix \eq{sqmon} this is a completely smooth limit. However, limits of this kind often change the configuration of lines of the Baxter lattice in a radical way, and in fact in our case we found that the resulting graph cannot be drawn at all on a conventional loom. However, it can be drawn once we allow the generalisation discussed in section \ref{sec:gen} when we can place internal vertices inside open faces. This gives the square graph shown earlier on figure \ref{fig:sqex1}, and on figure \ref{fig:sqlab} we have added the labels for $\Delta$'s corresponding to our notation here. 

From  \eq{sdrel1} we see that a convenient choice is to set two of the parameters to be
\beq
\label{d53}
    \Delta_5=D-\Delta_2\ ,  \ \ \ \Delta_3=D-\Delta_1
\eeq
which gives $\Delta_6=\Delta_4=0$. Then we are left with three independent parameters $\Delta_1,\Delta_2,\Delta_{1'}$ through which the remaining dimensions are still expressed by \eq{sdrel2}. Notice that \eq{d53} are again examples of nonlocal relations between the dimensions for our graph, as they do not follow from just demanding the sum of dimensions at each vertex to be $D$.

We can smoothly implement the limit $\Delta_4,\Delta_6\to 0$ directly for the monodromy matrix \eq{sqmon} of the original 6-point graph. We notice that the labels of Lax operators $L_4$ and $L_6$, which correspond to legs we have removed, become such that they are immediately diagonalised due to \eq{laxon1} (explicitly, they become $L_6[\Delta_{(11')}+D/2,\Delta_{(11')}+D]$ and $L_4[D,3D/2]$).
Therefore they can be simply removed from the monodromy matrix and we are left with a monodromy matrix acting now only on the four legs of our graph. It has the form
\beq
    L_5[\Delta_{(11')}+D/2,\Delta_{(121')}]L_3[D,\Delta_1+D/2]L_2[\Delta_{(121')}-D/2,\Delta_{(11')}]L_1[\Delta_1,D/2] \ .
\eeq
Then we find the evaluation parameters to be (setting $v_1=0$ )
\beq
    v_k=\left\{0,-\Delta _1'-\frac{\Delta _1}{2}-\frac{\Delta _2}{2}+D/2,-D/2,-\Delta _1'-\frac{\Delta _1}{2}-\frac{\Delta _2}{2}\right\} \ .
\eeq

Plugging them into \eq{vres}, \eq{ph} gives the differential equation satisifed by this graph. Let us mention that this graph (albeit with some propagators being massive) was partially discussed in \cite{Loebbert:2020hxk} and now we are able to provide a clear criterion for its Yangian invariance in the massless case, namely the relations \eq{d53}, \eq{sdrel2} between the scaling dimensions\footnote{In any case, however, as long as this graph is conformal it reduces to just a 2-point integral by application of star-triangle identity on two opposite vertices}.

Let us finally also mention that both for the square with 4 legs and the one with 6 legs we have checked Yangian invariance explicitly by repeated application of the intertwining relation and other properties of the Lax operators, serving as a nontrivial test of the general procedure for removing the lasso that we described in section \ref{sec:yang}.

\section{Yangian invariance with infinite-dimensional auxiliary space}

\label{sec:infd}

In this section we show that the Feynman graphs we discuss are also invariant under the action of the monodromy matrix that has an infinite-dimensional auxiliary space. That is,  we take the auxiliary space to be a representation of the same type as the physical one, labelled by a scaling dimension $\Delta_a$ which is now an extra parameter in the construction. This representation cannot be obtained by fusion from finite-dimensional representations and thus its application to the Yangian symmetry potentially provides new and  potentially powerful constraints on Feynman graphs. 
While the case we discussed above led to differential equations for the graphs, here we will get integral equations as the corresponding R-matrix is now an integral operator.  Let us notice that this construction is new even for the simplest $D=4$ fishnet with a quartic interaction and it would be interesting to explore its implications. At the same time,  those new  Yangian integral equations represent new relations establishing equivalence between very different  Feynman diagrams.

\begin{figure}[h]
 \centering
   \includegraphics[scale=1]{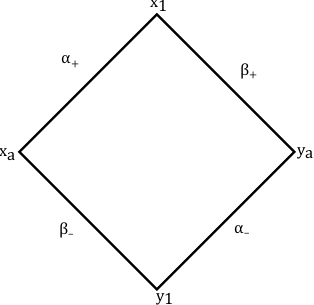}  \caption{The kernel of the R-matrix as an integral operator. The physical space corresponds to the vertical direction and the auxiliary space to the horizontal direction.}
   \label{fig:rker}
 \end{figure}

The conformal R-matrix for the case we are discussing was constructed in \cite{Chicherin:2012yn}\footnote{our notation differs from that paper by $R_{12}^{\rm here}=\left.R_{12}^{\rm there}P_{12}\right|_{\Delta_1\leftrightarrow\Delta_2}$; we mostly follow the notation of the review \cite{ferrando:tel-03987820}}. Let us label the physical and auxiliary spaces as '$1$' and '$a$' respectively, with the corresponding scaling dimensions
\beq
    \Delta_1=\Delta_{phys} \ , \ \Delta_a=\Delta_{aux} \ .
\eeq
The tensor product of these two spaces corresponds to the space of functions $f(x_1,x_a)$. Then the kernel of the R-matrix is given on figure \ref{fig:rker} and it acts on a function as
\beq
\label{r12}
    (R_{1a}f)(x_1,x_a)=
    4^{2(u-v)}A(-(u-v)-\rho)A(-(u-v)+\rho)
    \int \frac{d^Dy_1d^Dy_a\;f(y_1,y_a)}{x_{1a}^{2\alpha_+}(x_1-y_a)^{2\beta_+}(x_a-y_1)^{2\beta_-}y_{1a}^{2\alpha_-}} \ ,
\eeq
where we denoted
\beq
\label{au}
    \alpha_+=v_+ - u_- \ , \ \alpha_- = v_- - u_+\ , \ 
    \beta_-=u_+-v_++D/2 \ , \ 
    \beta_+=u_--v_-+D/2
\eeq
and
\beq
\label{ud}
    u_+=u+\frac{\Delta_a-D}{2} \ , \ u_-=u-\frac{\Delta_a}{2} \ , \
    v_+=v+\frac{\Delta_1-D}{2} \ , \
    v_-=v-\frac{\Delta_1}{2} \ .
\eeq
We also introduced
\beq
    \rho=\frac{\Delta_1-\Delta_a}{2}
\eeq
and
\beq
    A(u)=\frac{\Gamma(D/2-u)}{\Gamma(u)} \ .
\eeq
In our notation the first space is the physical one (vertical direction on the figure) and the second one is auxiliary (horizontal direction).

Notice that we have four labels $\alpha_\pm,\beta_\pm$ but the R-matrix actually depends only on three parameters: $\Delta_1,\Delta_a$ and the spectral parameter $u-v$. However the labels defined in \eq{au} automatically satisfy an additional constraint  
\beq
    \alpha_+ + \beta_+ + \alpha_- + \beta_- = D
\eeq
ensuring the matching of the number of parameters.
Notice also that the individual labels, unlike the final R-matrix, depend separately on $u$ and $v$ so the notation is somewhat redundant.

\subsection{Chain and cross relations}

\begin{figure}[h]
 \centering
   \includegraphics[scale=1]{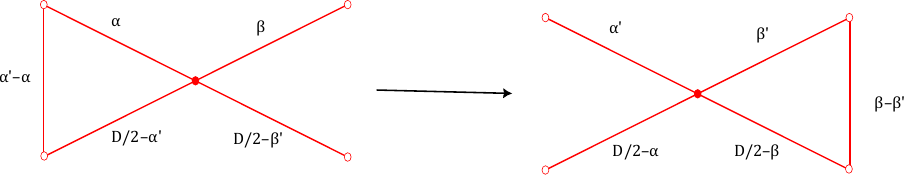}  \caption{The cross relation which holds for $\alpha+\beta=\alpha'+\beta'$.}
   \label{fig:crossrel}
 \end{figure}

Below we will make use of several important identities which allow us to transform expressions built from these R-matrices. First, we will use the chain relation which reads
\beq
    \int d^Dz\frac{1}{(x-z)^{2\alpha}(z-y)^{2\beta}}=\frac{\pi^{D/2}A(\alpha)A(\beta)A(D-\alpha-\beta)}{(x-y)^{2(\alpha+\beta-D/2)}} \ .
\eeq
One can also derive from it a representation of the delta-function in terms of a propagator,
\beq
\label{deps}
    \lim_{\epsilon\to 0}\frac{\epsilon}{(x-y)^{2(D/2-\epsilon)}}=\frac{\pi^{D/2}}{\Gamma(D/2)}\delta(x-y) \ .
\eeq
Second, we will use the cross  relation shown on figure \ref{fig:crossrel} which is a consequence of the star-triangle identity and allows us to move a propagator through a quartic integration point. It reads
\beqa &&
\label{cr1}
\int \frac{d^Dz}{|x_1-z|^{2(D/2-\alpha')}|x_2-z|^{2\alpha}|y_1-z|^{2(D/2-\beta')}|y_2-z|^{2\beta}} 
\\ \nn &&
=\int \frac{d^Dz}{|x_1-z|^{2\alpha'}|x_2-z|^{2(D/2-\alpha)}|y_1-z|^{2\beta'}|y_2-z|^{2(D/2-\beta)}}
\\ \nn &&
\ \ \ \times A(\alpha)A(\beta)A(D/2-\alpha')A(D/2-\beta')
\eeqa
and is satisfied as long as
\beq
\label{aap}
\alpha+\beta=\alpha'+\beta' \ .
\eeq

\subsection{Properties of the R-matrix}

Our main idea is to show that analogs of the key properties \eq{intw}, \eq{laxon1}, \eq{lte} of the Lax operator with finite-dimensional auxiliary space we used above have direct analogs for the infinite-dimensional case, i.e. for the R-matrix \eq{r12}. After that the whole lasso construction from section \ref{sec:yang} can be used without changes and will ensure the invariance of the Feynman graphs.

\subsubsection{Intertwining relation}

\begin{figure}[h]
 \centering
   \includegraphics[scale=0.53]{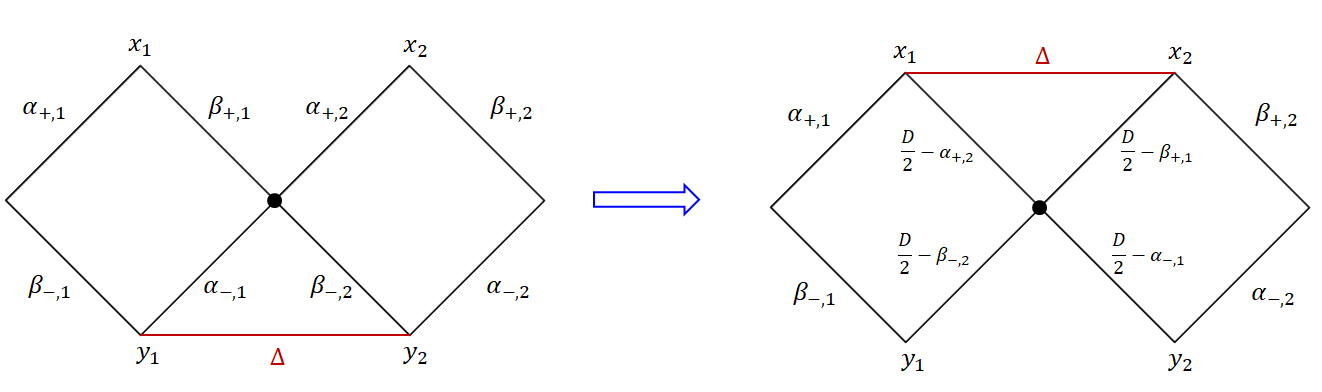}  \caption{Moving the propagator through two R-matrices.}
   \label{fig:crr}
 \end{figure}

The first key property we will need is an analog of the intertwining relation \eq{intw} which allows one to move a propagator through a product of two R-matrices. Remarkably it also extends to our case and can be derived  using the cross relation \eq{cr1}. Concretely, let us consider two R-matrices $R_{1a}$ and $R_{2a}$, with $R_{ka}$ acting on the $k$-th physical space and in the auxiliary space. We will introduce the following convenient notation for the R-matrices. First, denote by $R(u_+,u_-,v_+,v_-)$ the R-matrix defined as above by \eq{r12} with conventions \eq{au}, \eq{ud}. To write the intertwining relation in a way similar to the finite-dimensional case \eq{intw}, let us introduce the operator
\beq
\label{lidef}
    L^{\rm inf}_{k}(w_+,w_-)=R_{ka}((\Delta_a-D)/2,-\Delta_a/2,-D/2-w_-,-D/2-w_+)
\eeq
which acts in the tensor product of the $k$-th physical space and the auxiliary space. Here the arguments are chosen in such a way that the scaling dimension for the auxiliary space is $\Delta_a$ while the two parameters $w_\pm$ are related to the R-matrix  spectral parameter\footnote{The arguments in \eq{lidef} are chosen so that in \eq{ud} we have $u=0$ and the spectral parameter of the R-matrix in those conventions is $-v$.} $u-v\equiv u_R$ and the scaling dimension in the physical space by exactly the same relations \eq{du} as in the finite-dimensional case, namely
 \beq
    u_R=\frac{1}{2}(w_+ + w_- +D/2) \ , \ \ \Delta_{phys}=w_+-w_-+D/2 \ .
 \eeq
The intertwining relation then takes the form 
\beq
\label{lintw}
     L^{\rm inf}_{1}(z+\Delta,w) L^{\rm inf}_{2}(w',z)x_{12}^{-2\Delta}=
 F(z,\Delta,\Delta_a)x_{12}^{-2\Delta}L^{\rm inf}_{1}(z,w)L^{\rm inf}_{2}(w',z+\Delta)
\eeq
where
\beq
\label{fdef}
  F(z,\Delta,\Delta_a)=
     A\left(-\Delta -\frac{\Delta _a}{2}-z\right) A\left(D+\Delta -\frac{\Delta _a}{2}+z\right)
   A\left(-\frac{D}{2}+\frac{\Delta _a}{2}-z\right) A\left(\frac{D}{2}+\frac{\Delta
   _a}{2}+z\right) \ .
\eeq
In order to derive it, one considers the lhs shown graphically on figure \ref{fig:crr} and uses the cross relation \eq{cr1} to move the propagator (shown in red) through the quartic integration point in the middle. This leads to a change of powers of the various propagators, and one also needs to ensure that the cross relation can be applied at all (i.e. the condition \eq{aap} is satisfied). Taking all this into account and keeping careful track of the notation, we find as result of a somewhat tedious calculation that we get the simple relation \eq{lintw}. Notice that, in particular, applicability of the cross relation restricts both R-matrices to be constructed with the same value of the scaling dimension $\Delta_a$ for the auxiliary space. We can see, in the notation we have chosen, that this relation has exactly the same form as its counterpart \eq{intw} for the finite-dimensional case!

\subsubsection{Diagonal action on constants}

The second key property we will use is an analog of \eq{laxon1} and \eq{lte}, i.e. a simple action on constant functions when the Lax operator arguments are adjusted in a special way corresponding to $\Delta_1=0$ or $\Delta_1=D$. Consider first the action of $R_{1a}$ on a function $f=f(x_a)$ (independent of $x_1$) when $\Delta_1\to 0$. It gives, using the chain relation for $y_1$ integration and then using \eq{deps},
\beqa
    (R_{1a}f)=&&
    \pi^{D/2}4^{2(u-v)}A(-(u-v)-\rho)A(-(u-v)+\rho)A(\alpha_-)A(\beta_-)
    \times
    \\ \nn &&
   \frac{\Gamma(D/2-\Delta_1)}{\Gamma(\Delta_1)} \frac{1}{x_{1a}^{2\alpha_+}}\int \frac{d^Dy_a\;f(y_a)}{(x_1-y_a)^{2\beta_+}(x_a-y_a)^{2(D/2-\Delta_1)}}
        \\ \nn 
    = && 
    \pi^{D}4^{2(u-v)}A(-(u-v)+\Delta_a/2)A(-(u-v)-\Delta_a/2+D/2)
    f(x_a) \ .
\eeqa
To write this in a way that looks similar to the finite-dimensional case \eq{laxon1}, let us take $f$ to be a delta-function $\delta(x-y)$, then we can use $x,y$ as an analog of the matrix indices in \eq{laxon1}\footnote{To be precise, for an integral operator such as the R-matrix acting in the tensor product of the $k$-th physical space and the auxiliary space as an integral operator with kernel $R(x_k,x_a|y_k,y_a)$ so that $Rf(x_k,x_a)=\int dy_kdy_a R(x_k,x_a|y_k,y_a)f(y_k,y_a)$, we define its 'matrix element' with indices $x,y$ as an operator that acts on functions in the physical space as $R_{xy}f(x_k)=\int dy_k R(x_k,x|y_k,y)f(y_k)$.}. Then using also the notation \eq{lidef} we find
\beq
\label{li1}
     L^{\rm inf}(u,u+D/2)_{xy
     }\cdot 1=    G(u,\Delta_a)\delta(x-y)
\eeq
where we defined
\beq
\label{gdef}
    G(u,\Delta_a)=(4\pi)^{D}4^{2u}A(-u-D/2+\Delta_a/2)A(-u-\Delta_a/2) \ .
\eeq
This relation looks exactly like \eq{laxon1} up to the prefactor $G$.

Furthermore, as before we will also use the R-matrix 'transposed' in the physical space, defined by the property (schematically)
\beq
    \int dx f(x) (Rg(x))=\int dx (R^Tf(x))g(x) \ .
\eeq
Its kernel is obtained from the original one in \eq{r12} by simply exchanging $x_1\leftrightarrow y_1$. 
Similarly to the finite-dimensional case, here we find for it
\beq
\label{lit1}
   L^{\rm inf,T}(u+D/2,u)_{xy}\cdot 1=    G(u,\Delta_a)\delta(x-y) \ .
\eeq
The two properties \eq{li1} and \eq{lit1} are direct analogs of the ones we had for the finite-dimensional case (\eq{laxon1}, \eq{lte}) and in the notation we have chosen they are nicely written in the same way.

\subsection{Monodromy matrix and examples}

Above we have shown that direct counterparts of relations \eq{intw}, \eq{laxon1}, \eq{lte} we had for the finite-dimensional case exist as well for the R-matrices with infinite-dimensional auxiliary space in the form \eq{lintw}, \eq{li1}, \eq{lit1}. This means that the whole lasso construction from section \ref{sec:yang} goes through and leads to invariance of Feynman graphs under the action of the monodromy matrix constructed as 
\beq
    M_{xy}=(L_n^{\rm inf}(u+\delta_n^+,u+\delta_n^-)\dots L_1^{\rm inf}(u+\delta_1^+,u+\delta_1^-))_{xy}
\eeq
where $n$ is the number of external legs of the graph and the labels $\delta_k^\pm$ are chosen according to the same rules as discussed in section \ref{sec:yang}. Thus we have
\beq
\label{Mi}
    M_{xy}\ket{{\rm graph}}=\lambda(u)\delta(x-y)\ket{{\rm graph}}
\eeq
where the eigenvalue $\lambda(u)$ can be found by applying step by step the procedure of removing the lasso from the diagram discussed in section \ref{sec:yang}. In practice when computing the eigenvalue one should pay attention to the extra factors $F$ and $G$ in the intertwining relation \eq{lintw} and the diagonal action \eq{li1}, \eq{lit1} that were not present or very simple in the original construction. Notice also several nontrivial differences here compared to the finite-dimensional case:
\begin{itemize}
\item The invariance condition for the Feynman graph is now an integral rather than a differential equation
\item Since the R-matrices are built out of propagators, the lhs of the eigenvalue equation \eq{Mi} is itself a Feynman graph, thus this equation can be viewed as a relation between two different Feynman graphs.
\item Instead of discrete indices $\alpha,\beta$ in \eq{gre} here we have continuous labels $x,y$
\item We have an extra parameter in the monodromy matrix, namely the scaling dimension $\Delta_a$ associated to the auxiliary space
\end{itemize}

All these features look rather intriguing and we hope they should lead to new constraints for Feynman graphs. While we postpone a more detailed investigation of the construction to the future, below we will illustrate it on the example of the cross and double cross integrals.

\subsubsection{Example: cross integral}

\begin{figure}[h]
 \centering
   \includegraphics[scale=0.5]{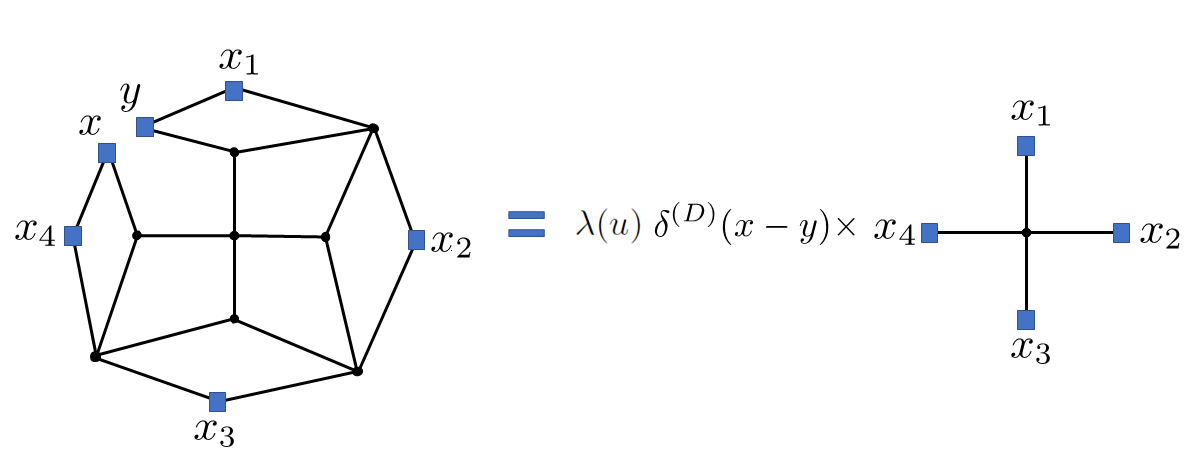}  \caption{Action on the cross integral of the Yangian monodromy matrix built from $R$-matrices with non-compact representation in auxiliary space. The cross integral is an eigenfunction of this monodromy matrix.}
   \label{fig:crm}
 \end{figure}

Let us first consider the 4-point cross integral \eq{i41}. 
For simplicity let us focus on the case when the dimension $D$ is arbitrary but all four propagator scaling dimensions are set to $D/4$. Then repeating the steps from section \ref{sec:cross}, we find that the monodromy matrix in this case is
\beqa
    M= L_4^{\rm inf}[D,5D/4]
    L_3^{\rm inf}[3D/4,D]
      L_2^{\rm inf}[D/2,3D/4]
    L_1^{\rm inf}[D/4,D/2]
\eeqa
where we used the notation
\beq
    L_k^{\rm inf}[\delta^+,\delta^-]=L_k^{\rm inf}(u+\delta^+,u+\delta^-) \ .
\eeq
We show the action of the monodromy matrix on the graph on figure \ref{fig:crm}.

\begin{figure}[h]
 \centering
   \includegraphics[scale=0.4]{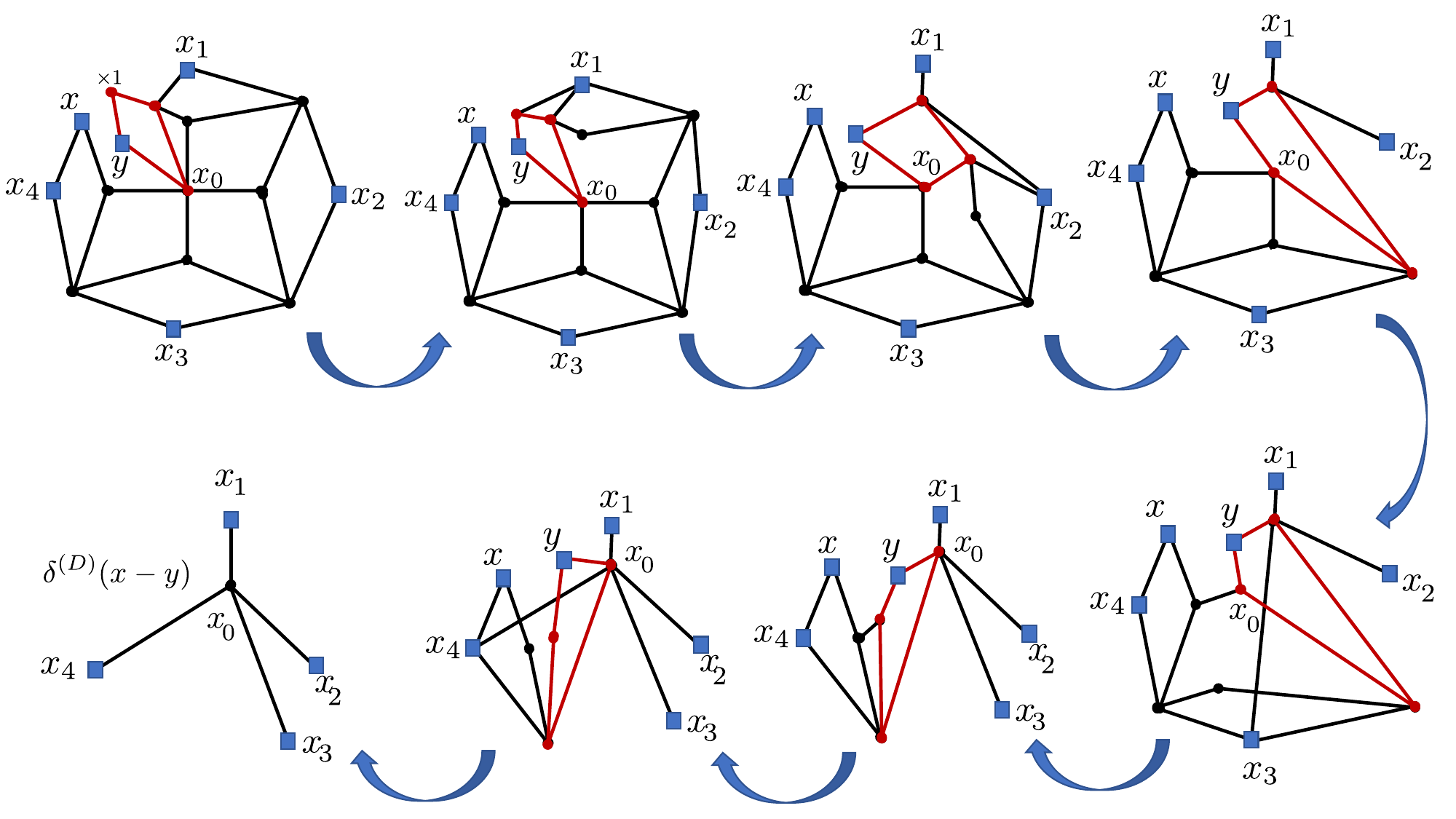}  \caption{We remove the parts of the 'lasso' acting on the cross integral one by one. As a result we find that the integral is an eigenstate of the monodromy matrix. At the first step, we insert an extra $L_0$ (drawn in red) using the identity \eqref{lit1}, then we apply the cross relation (figure \ref{fig:crossrel}) to move  outside the propagator coming out of the point $x_0$ vertically, which leads to the 2nd configuration; applying the chain relation  to the point where only two propagators meet,  then the cross relation moving outside the right propagator stemming from the point $x_0$ and again the identities \eqref{li1}, we come to the 3rd configuration, etc. At the last step, we apply two chain relations and, again due to \eqref{li1}, we restore the original cross graph, thus proving the Yangian invariance equation for it.    }
   \label{fig:cri}
 \end{figure}

In order to compute the eigenvalue, like before we introduce an extra operator $L_0^{{\rm inf},T}$ acting on the integration coordinate $x_0$ and then move the propagators one by one through the L-operators, repeating the steps in section \ref{sec:cross}. We give a schematic representation of this process on figure \ref{fig:cri}. 
 Collecting all the $F$ and $G$ factors originating from \eq{lintw}, \eq{li1} we find that the eigenvalue reads
\beq
\label{eigc}
   \lambda(u)=F(u)G\left(u+\frac{D}{2}\right)\prod_{k=1}^{3}F\left(u+\frac {kD}{4}\right)G\left(u+\frac {kD}{4}\right)
\eeq
where $F$ and $G$ correspond to the definitions \eq{fdef}, \eq{gdef} with $\Delta=D/4$ and $\Delta_a$ kept arbitrary.
For the case when $D=4$ (with $\Delta=1$) this can be simplified to
\beqa
\nn\lambda(u)=&&
    \frac{\left(-\frac{\Delta _a}{2}+u+4\right) \left(-\frac{\Delta _a}{2}+u+5\right){}^2
   \left(\frac{\Delta _a}{2}+u+1\right) \left(\frac{\Delta _a}{2}+u+2\right) \left(\frac{\Delta
   _a}{2}+u+5\right)}{\left(-\frac{\Delta _a}{2}+u+3\right) \left(-\frac{\Delta _a}{2}+u+6\right)
   \left(-\frac{\Delta _a}{2}+u+7\right) \left(\frac{\Delta _a}{2}+u+3\right){}^2
   \left(\frac{\Delta _a}{2}+u+4\right)}
   \\  && \times \pi^{16}4^{8u+32}
   \left(\frac{A(\Delta_a/2-u-3)}{A(\Delta_a/2+u+5)}\right)^4 \ .
\eeqa

\subsubsection{Example: double cross integral}

\begin{figure}[t]
 \centering
   \includegraphics[scale=1]{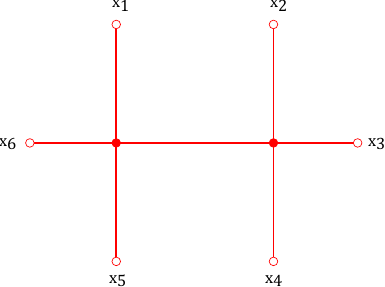}  \caption{The double cross integral.}
   \label{fig:dc}
 \end{figure}

As another example, consider the more involved 6-point double cross integral, shown on figure \ref{fig:dc}. Again we will for simplicity take all propagator scaling dimensions to be $D/4$ (generalization to the generic case is straightforward). Then we find that the monodromy matrix whose eigenstate it is has the form
\beq
    L_6^{\rm inf}[D,5D/4]L_5^{\rm inf}[3D/4,D]L_4^{\rm inf}[3D/4,D]L_3^{\rm inf}[D/2,3D/4]L_2^{\rm inf}[D/4,D/2]L_1^{\rm inf}[D/4,D/2] \ .
\eeq
To compute the eigenvalue we again introduce two new operators $L_0^{\rm inf}$ and $L_{0'}^{\rm inf}$ at the two integration points and  commute the propagators through all the L-operators, via a calculation similar to the one discussed for the double cross (for the usual fishnet theory and usual Lax operators with 4d auxiliary space) in \cite{Chicherin:2017cns}. In the end we find the eigenvalue to be (in the same notation as in \eq{eigc})
\beq
    \lambda(u)=
    F(u)^2G\left(u+\frac{D}{2}\right)\prod_{k=1}^3\left[F\left(u+\frac{kD}{4}\right)G\left(u+\frac{kD}{4}\right)\right]^{2-\delta_{k,3}} \ .
\eeq

The conformal double cross integral has been computed with generic dimensions but has a highly complicated form with 9-fold nested sums \cite{Loebbert:2019vcj,Ananthanarayan:2020ncn}. It would be interesting to see if the constraints coming from our integral Yangian invariance equation could help to simplify it.

\section{Conclusions}

\label{sec:concl}

In this work we derived new Yangian symmetry relations for integrable conformal planar Feynman graphs with disc topology, introduced by A.~Zamolodchikov~\cite{Zamolodchikov:1980mb}. Such graphs are ubiquitous for single trace correlators in generalized fishnet CFTs~\cite{Kazakov:2022dbd} in the planar 't~Hooft limit. Our Yangian relations generalize  those found in~\cite{Chicherin:2017cns,Chicherin:2017frs} for particular case of graphs with regular square lattice structure, and then in \cite{Loebbert:2019vcj,
Loebbert:2020hxk,Loebbert:2020tje,Loebbert:2020glj,Corcoran:2020epz,Corcoran:2021gda} for specific other cases, to the whole variety of such integrable graphs described  in~\cite{Zamolodchikov:1980mb}. The integrablity of these graphs was established in~\cite{Zamolodchikov:1980mb} via the star-triangle relations.  Our Yangian relations represent another manifestation of this integrability: the graph turns out to be an eigenfunction of the `lasso operator' acting on its external legs. This lasso operator is a monodromy matrix built out of  Lax operators with the spectral parameter changing according to the weights of external propagators. Its auxiliary space could be, a priori, in any representation of the underlying algebra. We consider here, apart from the standard compact representation, also the non-compact principal series representation of $D$-dimensional conformal algebra. Interestingly, this seems to produce some new relations connecting various conformal Feynman graphs as well as new integral equations for the graphs.

Let us point out a few  future  directions:
\begin{itemize}
    \item Using our new integral and differential equations it would be interesting to try to bootstrap various new Feynman integrals, extending the Yangian bootstrap program that has already brought novel results \cite{Loebbert:2019vcj,Ananthanarayan:2020ncn, Corcoran:2020epz,Loebbert:2022nfu}.
    \item One should study in more detail the implications of our integral equations for the Feynman graphs, in particular for the usual 4d fishnets as well as for 2d fishnets related to Calabi-Yau geometry \cite{intinprog}. In the latter case known Yangian constraints provide Picard-Fuchs differential equations for periods of the CY manifold, whereas our construction may give some constraints of a different type whose geometric role would be interesting to uncover.
    \item It is important to  explore the interplay between the Yangian symmetry for correlators and modern separation of variables (SoV) methods. The latter have seen remarkable progress recently  for spin chain correlators \cite{Cavaglia:2019pow,Gromov:2019wmz,Maillet:2020ykb,Gromov:2020fwh,Gromov:2022waj} and are starting to be used for computation of correlation functions in $N=4$ SYM as well \cite{Cavaglia:2018lxi,Giombi:2018qox,Bercini:2022jxo}, and for which the groundwork in the fishnet case laid down  in \cite{Cavaglia:2021mft}.
    \item  A related question is to derive the Quantum Spectral Curve/Baxter equations \cite{Gromov:2013pga,Gromov:2017cja} for the spectrum of the vast family of loom CFTs from \cite{Kazakov:2022dbd}, which should open the way to compute the spectrum in a wide variety of regimes as was done for the standard fishnet theory \cite{Gromov:2017cja,Cavaglia:2020hdb} (as well as for the parent $\gamma$-deformed super Yang-Mills theory \cite{Levkovich-Maslyuk:2020rlp,Kazakov:2015efa}), see \cite{Gurdogan:2020ppd} for recent results from diagrams in this context.
        \item It would be interesting to investigate the manifestation of Yangian invariance in the weak/strong dual model to the fishnet CFT known as the fishchain \cite{Gromov:2019aku,Gromov:2019bsj,Gromov:2019jfh,Gromov:2021ahm}, which may also help to derive it for a larger class of fishnet theories.
        
           \item  Another future direction is studying generalisations of the graphs we considered here to the massive case and extending to this situation the Yangian bootstrap methods developed for massive Feynman integrals in  \cite{Loebbert:2020glj,Loebbert:2020tje,Loebbert:2020hxk}.
    \item It should be possible to generalize our results to graphs that are not purely scalar and such as fermionic or spinning  spinning conformal diagrams, of the type considered in~\cite{Caetano:2016ydc,Chicherin:2017frs,Gromov:2018hut,Derkachov:2019tzo,Derkachov:2021ufp,Derkachov:2021rrf,Kazakov:2022dbd,Kazakov:2018gcy,Kazakov:2018hrh}.
    
\item  While we have worked with graphs of disc topology, it would be interesting to try and extend the lasso methods to higher topologies such as cylinder/pair of pants, with possible applications to computing wrapping corrections (wheel diagrams) and structure constants, as well as clarifying the (related) role of diagrams with double traces.
    \item  
The  Yangian equations of the kind we get here are designed for the study of a specific type of conformal Feynman diagrams. However, we think that they  may have a much wider spectrum of applications. In particular, there should exist similar lasso operators for more familiar statistical mechanical systems, such as the 8-vertex  model and its generalisations. The integrability of many of these models, usually having the round-the-face interaction for Boltzmann weights,  is in fact also based on  star-triangle relations for discrete spin variables (see \cite{Bazhanov:2022wdj} for recent developments in this direction).

\end{itemize}

\section*{Acknowledgements}

We thank B.~Basso, V.~Bazhanov, A.~Cavaglia, G.~Ferrando, A.~Garbali,  N.~Gromov, G.~Korchemsky, I.~Kostov, J.~Lamers,  F.~Loebbert, V.~Mangazeev,
A.~Molev, E.~Olivucci, V.~Pasquier, P.~Ryan, D.~Serban, S.~Sergeev, P.~Wiegmann for related discussions.   
Part of this work was carried out during the authors' (V.K. and F.L.-M.) stay at the NCCR SwissMAP workshop `Integrability in Condensed Matter Physics and QFT' (3rd to 12th of February 2023) which took place at the SwissMAP Research Station. These authors would like to thank the Swiss National Science Foundation, which funds SwissMAP (grant number 205607) and, in addition, supported the event via the grant
 IZSEZ0\_215085.  V.K. would like to thank the Sydney Mathematical Research Institute (SMRI), the Australian  National University  and the Melbourne University, where  a part of this project was realized, for the financial support and hospitality.

 \bibliographystyle{JHEP.bst}
 \bibliography{FishYB}

\providecommand{\href}[2]{#2}\begingroup\raggedright\begin{thebibliography}{10}

\bibitem{Faddeev:1996iy}
L.D.~Faddeev, \emph{{How algebraic Bethe ansatz works for integrable model}},
  in \emph{{Relativistic gravitation and gravitational radiation. Proceedings,
  School of Physics, Les Houches, France, September 26-October 6, 1995}},
  pp.~pp. 149--219, 1996 [\href{https://arxiv.org/abs/9605187}{{\ttfamily
  9605187}}].

\bibitem{molev2007yangians}
A.~Molev, \emph{Yangians and classical Lie algebras}, no.~143, American
  Mathematical Soc. (2007).

\bibitem{Bernard:1992ya}
D.~Bernard, \emph{{An Introduction to Yangian Symmetries}},
  \href{https://doi.org/10.1142/S0217979293003371}{\emph{Int. J. Mod. Phys. B}
  {\bfseries 7} (1993) 3517}
  [\href{https://arxiv.org/abs/hep-th/9211133}{{\ttfamily hep-th/9211133}}].

\bibitem{MacKay:2004tc}
N.J.~MacKay, \emph{{Introduction to Yangian symmetry in integrable field
  theory}}, \href{https://doi.org/10.1142/S0217751X05022317}{\emph{Int. J. Mod.
  Phys. A} {\bfseries 20} (2005) 7189}
  [\href{https://arxiv.org/abs/hep-th/0409183}{{\ttfamily hep-th/0409183}}].

\bibitem{Loebbert:2016cdm}
F.~Loebbert, \emph{{Lectures on Yangian Symmetry}},
  \href{https://doi.org/10.1088/1751-8113/49/32/323002}{\emph{J. Phys. A}
  {\bfseries 49} (2016) 323002}
  [\href{https://arxiv.org/abs/1606.02947}{{\ttfamily 1606.02947}}].

\bibitem{Drummond:2009fd}
J.M.~Drummond, J.M.~Henn and J.~Plefka, \emph{{Yangian symmetry of scattering
  amplitudes in N=4 super Yang-Mills theory}},
  \href{https://doi.org/10.1088/1126-6708/2009/05/046}{\emph{JHEP} {\bfseries
  05} (2009) 046} [\href{https://arxiv.org/abs/0902.2987}{{\ttfamily
  0902.2987}}].

\bibitem{Drummond:2008vq}
J.M.~Drummond, J.~Henn, G.P.~Korchemsky and E.~Sokatchev, \emph{{Dual
  superconformal symmetry of scattering amplitudes in N=4 super-Yang-Mills
  theory}}, \href{https://doi.org/10.1016/j.nuclphysb.2009.11.022}{\emph{Nucl.
  Phys. B} {\bfseries 828} (2010) 317}
  [\href{https://arxiv.org/abs/0807.1095}{{\ttfamily 0807.1095}}].

\bibitem{Beisert:2010gn}
N.~Beisert, J.~Henn, T.~McLoughlin and J.~Plefka, \emph{{One-Loop
  Superconformal and Yangian Symmetries of Scattering Amplitudes in N=4 Super
  Yang-Mills}}, \href{https://doi.org/10.1007/JHEP04(2010)085}{\emph{JHEP}
  {\bfseries 04} (2010) 085} [\href{https://arxiv.org/abs/1002.1733}{{\ttfamily
  1002.1733}}].

\bibitem{Arkani-Hamed:2012zlh}
N.~Arkani-Hamed, J.L.~Bourjaily, F.~Cachazo, A.B.~Goncharov, A.~Postnikov and
  J.~Trnka, \emph{{Grassmannian Geometry of Scattering Amplitudes}}, Cambridge
  University Press (4, 2016),
  \href{https://doi.org/10.1017/CBO9781316091548}{10.1017/CBO9781316091548},
  [\href{https://arxiv.org/abs/1212.5605}{{\ttfamily 1212.5605}}].

\bibitem{Huang:2010qy}
Y.-t.~Huang and A.E.~Lipstein, \emph{{Dual Superconformal Symmetry of N=6
  Chern-Simons Theory}},
  \href{https://doi.org/10.1007/JHEP11(2010)076}{\emph{JHEP} {\bfseries 11}
  (2010) 076} [\href{https://arxiv.org/abs/1008.0041}{{\ttfamily 1008.0041}}].

\bibitem{Bargheer:2010hn}
T.~Bargheer, F.~Loebbert and C.~Meneghelli, \emph{{Symmetries of Tree-level
  Scattering Amplitudes in N=6 Superconformal Chern-Simons Theory}},
  \href{https://doi.org/10.1103/PhysRevD.82.045016}{\emph{Phys. Rev. D}
  {\bfseries 82} (2010) 045016}
  [\href{https://arxiv.org/abs/1003.6120}{{\ttfamily 1003.6120}}].

\bibitem{Beisert:2017pnr}
N.~Beisert, A.~Garus and M.~Rosso, \emph{{Yangian Symmetry and Integrability of
  Planar N=4 Supersymmetric Yang-Mills Theory}},
  \href{https://doi.org/10.1103/PhysRevLett.118.141603}{\emph{Phys. Rev. Lett.}
  {\bfseries 118} (2017) 141603}
  [\href{https://arxiv.org/abs/1701.09162}{{\ttfamily 1701.09162}}].

\bibitem{Chicherin:2017cns}
D.~Chicherin, V.~Kazakov, F.~Loebbert, D.~Mueller and D.-l.~Zhong,
  \emph{{Yangian Symmetry for Bi-Scalar Loop Amplitudes}},
  \href{https://doi.org/10.1007/JHEP05(2018)003}{\emph{JHEP} {\bfseries 05}
  (2017) 003} [\href{https://arxiv.org/abs/1704.01967}{{\ttfamily
  1704.01967}}].

\bibitem{Chicherin:2017frs}
D.~Chicherin, V.~Kazakov, F.~Loebbert, D.~M\"uller and D.-l.~Zhong,
  \emph{{Yangian Symmetry for Fishnet Feynman Graphs}},
  \href{https://doi.org/10.1103/PhysRevD.96.121901}{\emph{Phys. Rev. D}
  {\bfseries 96} (2017) 121901}
  [\href{https://arxiv.org/abs/1708.00007}{{\ttfamily 1708.00007}}].

\bibitem{Chicherin:2013sqa}
D.~Chicherin and R.~Kirschner, \emph{{Yangian symmetric correlators}},
  \href{https://doi.org/10.1016/j.nuclphysb.2013.10.006}{\emph{Nucl. Phys. B}
  {\bfseries 877} (2013) 484}
  [\href{https://arxiv.org/abs/1306.0711}{{\ttfamily 1306.0711}}].

\bibitem{Duhr:2022pch}
C.~Duhr, A.~Klemm, F.~Loebbert, C.~Nega and F.~Porkert, \emph{Yangian-invariant
  fishnet integrals in 2 dimensions as volumes of calabi-yau varieties},
  \href{https://arxiv.org/abs/2209.05291}{{\ttfamily 2209.05291}}.

\bibitem{Loebbert:2019vcj}
F.~Loebbert, D.~M\"uller and H.~M\"unkler, \emph{{Yangian Bootstrap for
  Conformal Feynman Integrals}},
  \href{https://doi.org/10.1103/PhysRevD.101.066006}{\emph{Phys. Rev. D}
  {\bfseries 101} (2020) 066006}
  [\href{https://arxiv.org/abs/1912.05561}{{\ttfamily 1912.05561}}].

\bibitem{Loebbert:2020hxk}
F.~Loebbert, J.~Miczajka, D.~M\"uller and H.~M\"unkler, \emph{{Massive
  Conformal Symmetry and Integrability for Feynman Integrals}},
  \href{https://doi.org/10.1103/PhysRevLett.125.091602}{\emph{Phys. Rev. Lett.}
  {\bfseries 125} (2020) 091602}
  [\href{https://arxiv.org/abs/2005.01735}{{\ttfamily 2005.01735}}].

\bibitem{Loebbert:2020tje}
F.~Loebbert and J.~Miczajka, \emph{{Massive Fishnets}},
  \href{https://doi.org/10.1007/JHEP12(2020)197}{\emph{JHEP} {\bfseries 12}
  (2020) 197} [\href{https://arxiv.org/abs/2008.11739}{{\ttfamily
  2008.11739}}].

\bibitem{Loebbert:2020glj}
F.~Loebbert, J.~Miczajka, D.~M\"uller and H.~M\"unkler, \emph{{Yangian
  Bootstrap for Massive Feynman Integrals}},
  \href{https://doi.org/10.21468/SciPostPhys.11.1.010}{\emph{SciPost Phys.}
  {\bfseries 11} (2021) 010}
  [\href{https://arxiv.org/abs/2010.08552}{{\ttfamily 2010.08552}}].

\bibitem{Corcoran:2020epz}
L.~Corcoran, F.~Loebbert, J.~Miczajka and M.~Staudacher, \emph{{Minkowski Box
  from Yangian Bootstrap}},
  \href{https://doi.org/10.1007/JHEP04(2021)160}{\emph{JHEP} {\bfseries 04}
  (2021) 160} [\href{https://arxiv.org/abs/2012.07852}{{\ttfamily
  2012.07852}}].

\bibitem{Corcoran:2021gda}
L.~Corcoran, F.~Loebbert and J.~Miczajka, \emph{Yangian ward identities for
  fishnet four-point integrals},
  \href{https://doi.org/10.1007/JHEP04(2022)131}{\emph{JHEP} {\bfseries 04}
  (2022) 131} [\href{https://arxiv.org/abs/2112.06928}{{\ttfamily
  2112.06928}}].

\bibitem{Loebbert:2022nfu}
F.~Loebbert, \emph{{Integrability for Feynman Integrals}},  12, 2022
  [\href{https://arxiv.org/abs/2212.09636}{{\ttfamily 2212.09636}}].

\bibitem{Chicherin:2022nqq}
D.~Chicherin and G.P.~Korchemsky, \emph{{The SAGEX review on scattering
  amplitudes Chapter 9: Integrability of amplitudes in fishnet theories}},
  \href{https://doi.org/10.1088/1751-8121/ac8c72}{\emph{J. Phys. A} {\bfseries
  55} (2022) 443010} [\href{https://arxiv.org/abs/2203.13020}{{\ttfamily
  2203.13020}}].

\bibitem{Gurdogan:2015csr}
O.~G\"{u}rdo\u{g}an and V.~Kazakov, \emph{{New Integrable 4D Quantum Field
  Theories from Strongly Deformed Planar $\mathcal N = $ 4 Supersymmetric
  Yang-Mills Theory}}, \href{https://doi.org/10.1103/PhysRevLett.117.201602,
  10.1103/PhysRevLett.117.259903}{\emph{Phys. Rev. Lett.} {\bfseries 117}
  (2016) 201602} [\href{https://arxiv.org/abs/1512.06704}{{\ttfamily
  1512.06704}}].

\bibitem{Caetano:2016ydc}
J.a.~Caetano, O.~G\"urdo\u{g}an and V.~Kazakov, \emph{Chiral limit of
  $\mathcal{N}$ = 4 sym and abjm and integrable feynman graphs},
  \href{https://doi.org/10.1007/JHEP03(2018)077}{\emph{JHEP} {\bfseries 03}
  (2018) 077} [\href{https://arxiv.org/abs/1612.05895}{{\ttfamily
  1612.05895}}].

\bibitem{Kazakov:2018hrh}
V.~Kazakov, \emph{{Quantum Spectral Curve of $\gamma$-twisted ${\cal N}=4$ SYM
  theory and fishnet CFT}},  \href{https://arxiv.org/abs/1802.02160}{{\ttfamily
  1802.02160}}.

\bibitem{Kazakov:2018qbr}
V.~Kazakov and E.~Olivucci, \emph{{Biscalar Integrable Conformal Field Theories
  in Any Dimension}},
  \href{https://doi.org/10.1103/PhysRevLett.121.131601}{\emph{Phys. Rev. Lett.}
  {\bfseries 121} (2018) 131601}
  [\href{https://arxiv.org/abs/1801.09844}{{\ttfamily 1801.09844}}].

\bibitem{Mamroud:2017uyz}
O.~Mamroud and G.~Torrents, \emph{{RG stability of integrable fishnet models}},
  \href{https://doi.org/10.1007/JHEP06(2017)012}{\emph{JHEP} {\bfseries 06}
  (2017) 012} [\href{https://arxiv.org/abs/1703.04152}{{\ttfamily
  1703.04152}}].

\bibitem{Kazakov:2022dbd}
V.~Kazakov and E.~Olivucci, \emph{{The Loom for General Fishnet CFTs}},
  \href{https://arxiv.org/abs/2212.09732}{{\ttfamily 2212.09732}}.

\bibitem{Zamolodchikov:1980mb}
A.B.~Zamolodchikov, \emph{{'Fishnet' diagrams as a completely integrable
  system}}, \href{https://doi.org/10.1016/0370-2693(80)90547-X}{\emph{Phys.
  Lett.} {\bfseries 97B} (1980) 63}.

\bibitem{Derkachov:2021ufp}
S.~Derkachov, G.~Ferrando and E.~Olivucci, \emph{{Mirror channel eigenvectors
  of the d-dimensional fishnets}},
  \href{https://doi.org/10.1007/JHEP12(2021)174}{\emph{JHEP} {\bfseries 12}
  (2021) 174} [\href{https://arxiv.org/abs/2108.12620}{{\ttfamily
  2108.12620}}].

\bibitem{Derkachov:2021rrf}
S.~Derkachov and E.~Olivucci, \emph{{Conformal quantum mechanics \& the
  integrable spinning Fishnet}},
  \href{https://doi.org/10.1007/JHEP11(2021)060}{\emph{JHEP} {\bfseries 11}
  (2021) 060} [\href{https://arxiv.org/abs/2103.01940}{{\ttfamily
  2103.01940}}].

\bibitem{Derkachov:2020zvv}
S.~Derkachov and E.~Olivucci, \emph{{Exactly solvable single-trace four point
  correlators in $\chi$CFT$_4$}},
  \href{https://doi.org/10.1007/JHEP02(2021)146}{\emph{JHEP} {\bfseries 02}
  (2021) 146} [\href{https://arxiv.org/abs/2007.15049}{{\ttfamily
  2007.15049}}].

\bibitem{Derkachov:2019tzo}
S.~Derkachov and E.~Olivucci, \emph{{Exactly solvable magnet of conformal spins
  in four dimensions}},
  \href{https://doi.org/10.1103/PhysRevLett.125.031603}{\emph{Phys. Rev. Lett.}
  {\bfseries 125} (2020) 031603}
  [\href{https://arxiv.org/abs/1912.07588}{{\ttfamily 1912.07588}}].

\bibitem{Kazakov:2018gcy}
V.~Kazakov, E.~Olivucci and M.~Preti, \emph{{Generalized fishnets and exact
  four-point correlators in chiral CFT$_{4}$}},
  \href{https://doi.org/10.1007/JHEP06(2019)078}{\emph{JHEP} {\bfseries 06}
  (2019) 078} [\href{https://arxiv.org/abs/1901.00011}{{\ttfamily
  1901.00011}}].

\bibitem{Derkachov:2018rot}
S.~Derkachov, V.~Kazakov and E.~Olivucci, \emph{{Basso-Dixon Correlators in
  Two-Dimensional Fishnet CFT}},
  \href{https://arxiv.org/abs/1811.10623}{{\ttfamily 1811.10623}}.

\bibitem{Gromov:2018hut}
N.~Gromov, V.~Kazakov and G.~Korchemsky, \emph{{Exact Correlation Functions in
  Conformal Fishnet Theory}},
  \href{https://doi.org/10.1007/JHEP08(2019)123}{\emph{JHEP} {\bfseries 08}
  (2018) 123} [\href{https://arxiv.org/abs/1808.02688}{{\ttfamily
  1808.02688}}].

\bibitem{Gromov:2019aku}
N.~Gromov and A.~Sever, \emph{{Derivation of the Holographic Dual of a Planar
  Conformal Field Theory in 4D}},
  \href{https://doi.org/10.1103/PhysRevLett.123.081602}{\emph{Phys. Rev. Lett.}
  {\bfseries 123} (2019) 081602}
  [\href{https://arxiv.org/abs/1903.10508}{{\ttfamily 1903.10508}}].

\bibitem{Gromov:2017cja}
N.~Gromov, V.~Kazakov, G.~Korchemsky, S.~Negro and G.~Sizov,
  \emph{Integrability of conformal fishnet theory},
  \href{https://arxiv.org/abs/1706.04167}{{\ttfamily 1706.04167}}.

\bibitem{Grabner:2017pgm}
D.~Grabner, N.~Gromov, V.~Kazakov and G.~Korchemsky, \emph{{Strongly
  gamma-deformed N=4 SYM as an integrable CFT}},
  \href{https://doi.org/10.1103/PhysRevLett.120.111601}{\emph{Phys. Rev. Lett.}
  {\bfseries 28} (2017) e2476}
  [\href{https://arxiv.org/abs/1711.04786}{{\ttfamily 1711.04786}}].

\bibitem{Basso:2018cvy}
B.~Basso, J.~Caetano and T.~Fleury, \emph{{Hexagons and Correlators in the
  Fishnet Theory}}, \href{https://doi.org/10.1007/JHEP11(2019)172}{\emph{JHEP}
  {\bfseries 11} (2018) 172}
  [\href{https://arxiv.org/abs/1812.09794}{{\ttfamily 1812.09794}}].

\bibitem{Basso:2018agi}
B.~Basso and D.-l.~Zhong, \emph{{Continuum limit of fishnet graphs and AdS
  sigma model}}, \href{https://doi.org/10.1007/JHEP01(2019)002}{\emph{JHEP}
  {\bfseries 01} (2018) 002}
  [\href{https://arxiv.org/abs/1806.04105}{{\ttfamily 1806.04105}}].

\bibitem{Basso:2017jwq}
B.~Basso and L.J.~Dixon, \emph{{Gluing Ladder Feynman Diagrams into Fishnets}},
  \href{https://doi.org/10.1103/PhysRevLett.119.071601}{\emph{Phys. Rev. Lett.}
  {\bfseries 119} (2017) 071601}
  [\href{https://arxiv.org/abs/1705.03545}{{\ttfamily 1705.03545}}].

\bibitem{Cavaglia:2021mft}
A.~Cavagli\`a, N.~Gromov and F.~Levkovich-Maslyuk, \emph{{Separation of
  variables in AdS/CFT: functional approach for the fishnet CFT}},
  \href{https://doi.org/10.1007/JHEP06(2021)131}{\emph{JHEP} {\bfseries 06}
  (2021) 131} [\href{https://arxiv.org/abs/2103.15800}{{\ttfamily
  2103.15800}}].

\bibitem{Pittelli:2019ceq}
A.~Pittelli and M.~Preti, \emph{{Integrable fishnet from $\gamma$-deformed
  $\mathcal{N}=2$ quivers}},
  \href{https://doi.org/10.1016/j.physletb.2019.134971}{\emph{Phys. Lett. B}
  {\bfseries 798} (2019) 134971}
  [\href{https://arxiv.org/abs/1906.03680}{{\ttfamily 1906.03680}}].

\bibitem{Kostov:2022vup}
I.~Kostov, \emph{{Light-cone limits of large rectangular fishnets}},
  \href{https://doi.org/10.1007/JHEP03(2023)156}{\emph{JHEP} {\bfseries 03}
  (2023) 156} [\href{https://arxiv.org/abs/2211.15056}{{\ttfamily
  2211.15056}}].

\bibitem{Ferrando:2023ogg}
G.~Ferrando, A.~Sever, A.~Sharon and E.~Urisman, \emph{{A Large Twist Limit for
  Any Operator}},  \href{https://arxiv.org/abs/2303.08852}{{\ttfamily
  2303.08852}}.

\bibitem{baxter1978solvable}
R.J.~Baxter, \emph{Solvable eight-vertex model on an arbitrary planar lattice},
  {\emph{Philosophical Transactions of the Royal Society of London. Series A,
  Mathematical and Physical Sciences} {\bfseries 289} (1978) 315}.

\bibitem{Chicherin:2012yn}
D.~Chicherin, S.~Derkachov and A.P.~Isaev, \emph{{Conformal algebra: R-matrix
  and star-triangle relation}},
  \href{https://doi.org/10.1007/JHEP04(2013)020}{\emph{JHEP} {\bfseries 04}
  (2013) 020} [\href{https://arxiv.org/abs/1206.4150}{{\ttfamily 1206.4150}}].

\bibitem{Ananthanarayan:2020ncn}
B.~Ananthanarayan, S.~Banik, S.~Friot and S.~Ghosh, \emph{{Double box and
  hexagon conformal Feynman integrals}},
  \href{https://doi.org/10.1103/PhysRevD.102.091901}{\emph{Phys. Rev. D}
  {\bfseries 102} (2020) 091901}
  [\href{https://arxiv.org/abs/2007.08360}{{\ttfamily 2007.08360}}].

\bibitem{Derkachov:2001yn}
S.E.~Derkachov, G.P.~Korchemsky and A.N.~Manashov, \emph{{Noncompact Heisenberg
  spin magnets from high-energy QCD: 1. Baxter Q operator and separation of
  variables}}, \href{https://doi.org/10.1016/S0550-3213(01)00457-6}{\emph{Nucl.
  Phys.} {\bfseries B617} (2001) 375}
  [\href{https://arxiv.org/abs/hep-th/0107193}{{\ttfamily hep-th/0107193}}].

\bibitem{Derkachov:2002wz}
S.E.~Derkachov, G.P.~Korchemsky, J.~Kotanski and A.N.~Manashov,
  \emph{{Noncompact Heisenberg spin magnets from high-energy QCD. 2.
  Quantization conditions and energy spectrum}},
  \href{https://doi.org/10.1016/S0550-3213(02)00842-8}{\emph{Nucl. Phys. B}
  {\bfseries 645} (2002) 237}
  [\href{https://arxiv.org/abs/hep-th/0204124}{{\ttfamily hep-th/0204124}}].

\bibitem{Derkachov:2002tf}
S.E.~Derkachov, G.P.~Korchemsky and A.N.~Manashov, \emph{{Separation of
  variables for the quantum SL(2,R) spin chain}},
  \href{https://doi.org/10.1088/1126-6708/2003/07/047}{\emph{JHEP} {\bfseries
  07} (2003) 047} [\href{https://arxiv.org/abs/hep-th/0210216}{{\ttfamily
  hep-th/0210216}}].

\bibitem{Usyukina:1992jd}
N.I.~Usyukina and A.I.~Davydychev, \emph{{An Approach to the evaluation of
  three and four point ladder diagrams}},
  \href{https://doi.org/10.1016/0370-2693(93)91834-A}{\emph{Phys. Lett.}
  {\bfseries B298} (1993) 363}.

\bibitem{Boos:1990rg}
E.E.~Boos and A.I.~Davydychev, \emph{{A Method of evaluating massive Feynman
  integrals}}, \href{https://doi.org/10.1007/BF01016805}{\emph{Theor. Math.
  Phys.} {\bfseries 89} (1991) 1052}.

\bibitem{ferrando:tel-03987820}
G.~Ferrando, \emph{{Non-compact integrable spin chain for the conformal fishnet
  theory}}, theses, {Universit{\'e} Paris sciences et lettres}, Sept., 2021.

\bibitem{intinprog}
{Work in progress}.

\bibitem{Cavaglia:2019pow}
A.~Cavagli\`a, N.~Gromov and F.~Levkovich-Maslyuk, \emph{{Separation of
  variables and scalar products at any rank}},
  \href{https://doi.org/10.1007/JHEP09(2019)052}{\emph{JHEP} {\bfseries 09}
  (2019) 052} [\href{https://arxiv.org/abs/1907.03788}{{\ttfamily
  1907.03788}}].

\bibitem{Gromov:2019wmz}
N.~Gromov, F.~Levkovich-Maslyuk, P.~Ryan and D.~Volin, \emph{{Dual Separated
  Variables and Scalar Products}},
  \href{https://doi.org/10.1016/j.physletb.2020.135494}{\emph{Phys. Lett. B}
  {\bfseries 806} (2020) 135494}
  [\href{https://arxiv.org/abs/1910.13442}{{\ttfamily 1910.13442}}].

\bibitem{Maillet:2020ykb}
J.M.~Maillet, G.~Niccoli and L.~Vignoli, \emph{{On Scalar Products in Higher
  Rank Quantum Separation of Variables}},
  \href{https://doi.org/10.21468/SciPostPhys.9.6.086}{\emph{SciPost Phys.}
  {\bfseries 9} (2020) 086} [\href{https://arxiv.org/abs/2003.04281}{{\ttfamily
  2003.04281}}].

\bibitem{Gromov:2020fwh}
N.~Gromov, F.~Levkovich-Maslyuk and P.~Ryan, \emph{{Determinant form of
  correlators in high rank integrable spin chains via separation of
  variables}}, \href{https://doi.org/10.1007/JHEP05(2021)169}{\emph{JHEP}
  {\bfseries 05} (2021) 169}
  [\href{https://arxiv.org/abs/2011.08229}{{\ttfamily 2011.08229}}].

\bibitem{Gromov:2022waj}
N.~Gromov, N.~Primi and P.~Ryan, \emph{{Form-factors and complete basis of
  observables via separation of variables for higher rank spin chains}},
  \href{https://doi.org/10.1007/JHEP11(2022)039}{\emph{JHEP} {\bfseries 11}
  (2022) 039} [\href{https://arxiv.org/abs/2202.01591}{{\ttfamily
  2202.01591}}].

\bibitem{Cavaglia:2018lxi}
A.~Cavagli\`a, N.~Gromov and F.~Levkovich-Maslyuk, \emph{{Quantum spectral
  curve and structure constants in $ \mathcal{N}=4 $ SYM: cusps in the ladder
  limit}}, \href{https://doi.org/10.1007/JHEP10(2018)060}{\emph{JHEP}
  {\bfseries 10} (2018) 060}
  [\href{https://arxiv.org/abs/1802.04237}{{\ttfamily 1802.04237}}].

\bibitem{Giombi:2018qox}
S.~Giombi and S.~Komatsu, \emph{{Exact Correlators on the Wilson Loop in
  $\mathcal{N}=4$ SYM: Localization, Defect CFT, and Integrability}},
  \href{https://doi.org/10.1007/JHEP05(2018)109}{\emph{JHEP} {\bfseries 05}
  (2018) 109} [\href{https://arxiv.org/abs/1802.05201}{{\ttfamily
  1802.05201}}].

\bibitem{Bercini:2022jxo}
C.~Bercini, A.~Homrich and P.~Vieira, \emph{{Structure Constants in
  $\mathcal{N} = 4$ SYM and Separation of Variables}},
  \href{https://arxiv.org/abs/2210.04923}{{\ttfamily 2210.04923}}.

\bibitem{Gromov:2013pga}
N.~Gromov, V.~Kazakov, S.~Leurent and D.~Volin, \emph{{Quantum Spectral Curve
  for Planar $\mathcal{N} = 4$ Super-Yang-Mills Theory}},
  \href{https://doi.org/10.1103/PhysRevLett.112.011602}{\emph{Phys. Rev. Lett.}
  {\bfseries 112} (2014) 011602}
  [\href{https://arxiv.org/abs/1305.1939}{{\ttfamily 1305.1939}}].

\bibitem{Cavaglia:2020hdb}
A.~Cavaglia, D.~Grabner, N.~Gromov and A.~Sever, \emph{{Colour-twist operators.
  Part I. Spectrum and wave functions}},
  \href{https://doi.org/10.1007/JHEP06(2020)092}{\emph{JHEP} {\bfseries 06}
  (2020) 092} [\href{https://arxiv.org/abs/2001.07259}{{\ttfamily
  2001.07259}}].

\bibitem{Levkovich-Maslyuk:2020rlp}
F.~Levkovich-Maslyuk and M.~Preti, \emph{{Exploring the ground state spectrum
  of \ensuremath{\gamma}-deformed N = 4 SYM}},
  \href{https://doi.org/10.1007/JHEP06(2022)146}{\emph{JHEP} {\bfseries 06}
  (2022) 146} [\href{https://arxiv.org/abs/2003.05811}{{\ttfamily
  2003.05811}}].

\bibitem{Kazakov:2015efa}
V.~Kazakov, S.~Leurent and D.~Volin, \emph{{T-system on T-hook: Grassmannian
  Solution and Twisted Quantum Spectral Curve}},
  \href{https://doi.org/10.1007/JHEP12(2016)044}{\emph{JHEP} {\bfseries 12}
  (2015) 044} [\href{https://arxiv.org/abs/1510.02100}{{\ttfamily
  1510.02100}}].

\bibitem{Gurdogan:2020ppd}
O.~G\"urdo\u{g}an, \emph{{From integrability to the Galois coaction on Feynman
  periods}}, \href{https://doi.org/10.1103/PhysRevD.103.L081703}{\emph{Phys.
  Rev. D} {\bfseries 103} (2021) L081703}
  [\href{https://arxiv.org/abs/2011.04781}{{\ttfamily 2011.04781}}].

\bibitem{Gromov:2019bsj}
N.~Gromov and A.~Sever, \emph{{Quantum fishchain in AdS$_{5}$}},
  \href{https://doi.org/10.1007/JHEP10(2019)085}{\emph{JHEP} {\bfseries 10}
  (2019) 085} [\href{https://arxiv.org/abs/1907.01001}{{\ttfamily
  1907.01001}}].

\bibitem{Gromov:2019jfh}
N.~Gromov and A.~Sever, \emph{{The holographic dual of strongly
  $\gamma$-deformed $ \mathcal{N} $ = 4 SYM theory: derivation, generalization,
  integrability and discrete reparametrization symmetry}},
  \href{https://doi.org/10.1007/JHEP02(2020)035}{\emph{JHEP} {\bfseries 02}
  (2020) 035} [\href{https://arxiv.org/abs/1908.10379}{{\ttfamily
  1908.10379}}].

\bibitem{Gromov:2021ahm}
N.~Gromov, J.~Julius and N.~Primi, \emph{{Open fishchain in N = 4
  Supersymmetric Yang-Mills Theory}},
  \href{https://doi.org/10.1007/JHEP07(2021)127}{\emph{JHEP} {\bfseries 07}
  (2021) 127} [\href{https://arxiv.org/abs/2101.01232}{{\ttfamily
  2101.01232}}].

\bibitem{Bazhanov:2022wdj}
V.V.~Bazhanov and S.M.~Sergeev, \emph{{An Ising-type formulation of the
  six-vertex model}},
  \href{https://doi.org/10.1016/j.nuclphysb.2022.116055}{\emph{Nucl. Phys. B}
  {\bfseries 986} (2023) 116055}
  [\href{https://arxiv.org/abs/2205.10708}{{\ttfamily 2205.10708}}].

\end{thebibliography}\endgroup

\end{document}